\documentclass[a4paper,fleqn,usenatbib]{mnras}

\usepackage[T1]{fontenc}

\newcommand{\cont}{870\,$\mu$\text{m}}
\newcommand{\sersic}{S\'ersic}
\newcommand{\res}{0\farcs18}
\newcommand{\ar}{axis ratio}
\newcommand{\flux}{$S_{870}$}
\newcommand{\CII}{[C{\small II}]}
\newcommand{\photz}{$z_{\text{photo}}$}

\newcommand{\Reff}{$R_{\text{eff}}$}
\newcommand{\num}{153}
\usepackage{newtxtext,newtxmath}
\usepackage{caption}
\usepackage[T1]{fontenc}
\usepackage{ae,aecompl}

\usepackage{graphicx}	% Including figure files
\usepackage{amsmath}	% Advanced maths commands
\usepackage{amssymb}	% Extra maths symbols
\usepackage{longtable}

\title[AS2UDS:  dust continuum morphologies of SMGs]{An ALMA survey of
  the SCUBA-2 Cosmology Legacy Survey UKIDSS/UDS field:
  High-resolution dust continuum morphologies and the link between
  sub-millimetre galaxies and spheroid formation}

\author[Gullberg et al.]{B.\ Gullberg,$^{1}$\thanks{E-mail: bitten.gullberg@durham.ac.uk}
Ian Smail,$^{1}$
A.\ M.\ Swinbank,$^{1}$
U.\ Dudzevi\v{c}i\={u}t\.{e},$^{1}$ 
S.\ M.\ Stach,$^{1}$
\newauthor
A.\ P.\ Thomson,$^{2}$
O.\ Almaini, $^{3}$
C.\ C.\ Chen, $^{4}$
C.\ Conselice, $^{3}$
E.\ A.\ Cooke,$^{1}$
D.\ Farrah,$^{5,6}$ 
\newauthor
R.\ J.\ Ivison,$^{4,7}$
D.\ Maltby,$^{3}$
M.\ J.\ Micha{\l}owski,$^{8}$ 
J.\ M.\ Simpson,$^{9}$
D.\ Scott,$^{10}$ 
\newauthor
J.\ L.\ Wardlow,$^{1, 11}$
A.\ Weiss$^{12}$
\\
$^{1}$ Centre for Extragalactic Astronomy, Department of Physics, Durham University, South Road, Durham DH1 3LE, UK\\
$^{2}$ Jodrell Bank Centre for Astrophysics, The University of Manchester, Oxford Road, Manchester, M13 9PL, UK\\
$^{3}$ School of Physics and Astronomy, University of Nottingham, University Park, Nottingham, NG7 2RD, UK\\ 
$^{4}$ European Southern Observatory, Karl-Schwarzschild-Stra\ss e 2, D-85748 Garching bei M\"unchen, Germany\\
$^{5}$ Department of Physics and Astronomy, University of Hawaii, 2505 Correa Road, Honolulu, HI 96822, USA\\
$^{6}$ Institute for Astronomy, 2680 Woodlawn Drive, University of Hawaii, Honolulu, HI 96822, USA\\
$^{7}$ Institute for Astronomy, University of Edinburgh, Royal Observatory, Blackford Hill, Edinburgh EH9 3HJ, UK \\
$^{8}$ Astronomical Observatory Institute, Faculty of Physics, Adam Mickiewicz University, ul.\ Sloneczna 36, 60-286 Poznan, Poland\\
$^{9}$ Academia Sinica Institute of Astronomy and Astrophysics, No. 1, Sec. 4, Roosevelt Road, Taipei 10617, Taiwan\\
$^{10}$ Department of Physics and Astronomy, University of British Columbia, 6224 Agricultural Road, Vancouver, BC V6T 1Z1, Canada\\ 
$^{11}$ Department of Physics, Lancaster University, Lancaster, LA1 4YB, UK\\
$^{12}$ Max-Planck-Institut fur Radioastronomie, Auf dem Hugel 69 D-53121 Bonn, Germany
}

\date{Accepted XXX. Received YYY; in original form ZZZ}

\pubyear{2019}

\hypersetup{draft}
\begin{document}
\label{firstpage}
\pagerange{\pageref{firstpage}--\pageref{lastpage}}
\maketitle

\begin{abstract}
We present an analysis of the morphology and profiles of the dust
continuum emission in \num\ bright sub-millimetre galaxies (SMGs)
detected with ALMA at signal to noise ratios of $>8$ in
high-resolution \res\ ($\sim1$\,kpc)  \cont\ maps. 
We measure sizes, shapes and light profiles for the rest-frame far-infrared emission from these luminous star-forming systems and derive a 
median effective radius ($R_e$) of $0\farcs10\pm0\farcs04$ for our sample with a median flux of \flux$=5.6\pm0.2$\,mJy.
We find that the apparent axial ratio ($b/a$) distribution of the SMGs
peaks at $b/a\sim 0.63\pm 0.02 $ and is best described by triaxial morphologies, while their emission profiles are best fit by a \sersic\ model with $n \simeq 1.0\pm0.1$, similar to exponential discs. 
This combination of triaxiality and $n\sim1$ \sersic\ index are
characteristic of bars and we suggest that the bulk of the \cont\ dust
continuum emission in the central $\sim2$\,kpc of these galaxies arises from  bar-like structures. 
As such we caution against using the orientation of shape of the bright dust continuum emission at $\eqsim$ resolution to assess either the orientation of any disc on the sky or tits inclination.
By stacking our \cont\ maps we recover faint extended dust continuum emission on $\sim4$\,kpc scales which contributes $13\pm1$\% of the total \cont\ emission. 
The scale of this extended emission is similar to that seen for the molecular gas and rest-frame optical light in these systems, suggesting that it represents an extended dust and gas disc at radii larger than the more
active bar component. 
Including this component in our estimated size of the sources we derive a typical effective radius of $\simeq0\farcs15\pm0\farcs05$ or $1.2\pm0.4$\,kpc.
Our results suggest that kpc-scale bars are ubiquitous features of high star-formation rate systems at $z\gg1$,  while these systems also contain fainter and more extended gas and stellar envelopes.  We suggest that these features, seen some $10$--$12$\,Gyrs ago, represent the formation phase of the earliest galactic-scale components: stellar bulges.
 
\end{abstract}


%%%%%%%%%%%%%%%%%%%%%%%%%%%%%%%%%%%%%%%%%%%%%%%%%%

%%%%%%%%%%%%%%%%% BODY OF PAPER %%%%%%%%%%%%%%%%%%
%
%
%
\section{Introduction}
Sub-millimetre galaxies (SMGs) are a class of high-redshift dust obscured, but far-infrared luminous, galaxies with estimated star-formation rates of $\sim100$--$1000$\,M$_{\odot}$\,yr$^{-1}$ \citep{smail97, barger98, hughes98}. 

%ZZZ add references:
%barger'98 nature
%hughes'98 nature
%Dudzeviciute19 MNRAS submitted

The  high star-formation rates are similar to those measured for local ultra-luminous infrared galaxies (ULIRGs, e.g. \citealt{sanders96, tacconi08, engel10, riechers11, bothwell13}). The intense star-formation activity in local ULIRGs is believed to be triggered and fuelled by  mergers, resulting in irregular morphologies at UV/optical wavelengths, with single and double nuclei and tidal tails \citep[e.g.][]{clements96, farrah01,surace01, veilleux02, psychogyios16}.   Theoretical models
provide support for this suggestion:  hydrodynamical simulations of mergers  can result in   remnants with a central starburst event building a bulge. After $\gtrsim1$\,Gyr the merger remnant comprises a central bulge with in situ star formation and an extended disc/envelope dominated by stars formed before the merger \citep{hopkins13}. It has been similarly suggested that major and minor mergers  may also be the trigger for  the activity in the high-redshift SMG population \cite[e.g.][]{mcalpine19}.   

The spatial extents of local (U)LIRGs has been shown to vary strongly depending upon  the observed wavelength: with the highest star-formation rate (U)LIRGs displaying  the most extended emission in the optical, while at the same time showing the most compact emission in the mid-infrared, which is thought to trace the on-going star formation \citep{chen10, psychogyios16}. 
Optical depths effects are a likely explanation for these varying trends, and this suggest that the physical size measured in the optical is highly dependent on the geometry of the dust distribution \citep{calzetti07, psychogyios16}.   Comparisons of the rest-frame optical and far-infrared sizes of high-redshift SMGs have suggested similar behaviour, with much more extended
optical sizes, compared to those derived from interferometric observations in the sub-millimetre, which trace
the bulk of the star-formation activity visible in the rest-frame far-infrared waveband \citep{simpson15b, ikarashi15, hodge16, lutz16}. 

However, while there are similarities, there are also apparent
differences between the observed properties of SMGs and those of comparably strongly star-forming ULIRGs in the local Universe.   One notable difference being the large spatial extent of the star-formation activity in the high redshift sources, which was hinted at in early interferometric studies \citep{chapman04, sakamoto08, ivison12}.  This has now been clearly demonstrated by ALMA:  while the the typical extent of the starburst seen in local ULIRGs is of the order of a few 100's\,pc to $\sim1$\,kpc, the rest-frame far-infrared emission in high-redshift SMGs arises from a region  with an effective FWHM  of $\sim2$--$3$\,kpc \citep[e.g.][]{simpson15b,hodge16}.   There are also 
hints that the dust continuum morphologies of some high-redshift SMGs show features which are not found in the typically more complex local counterparts.  Thus recent high-resolution (0\farcs03--0\farcs3) studies with ALMA have found that the dust continuum emission in SMGs  has a  disc-like brightness profile   \citep{simpson15a,hodge16,ikarashi17,gullberg18}.   
While a study of six SMGs at $z\simeq2.5$ by \cite{hodge19} at
0\farcs07 resolution ($\sim0.5$\,kpc) has shown spatially resolved
\cont\ dust continuum morphologies with ``clump-like'' structures
bracketing elongated nuclear emission, reminiscent of  bars and rings \citep{kormendy13}. The  sizes of the ``bars'' and ``rings'' are in the ratio of $1.9\pm0.3$, consistent with that expected for Lindblad resonances.  If these are indeed bars and rings then analytic theory and numerical simulations \cite[e.g.][]{binney87, lynden-bell96,athanassoula03} have shown that the ring is formed by gas outside the point of co-rotation being driven outward, by angular momentum transfer, collect into a ring near the outer Lindblad resonance.  
At radii inside the point of co-rotation, however, the gas falls inwards to the centre creating the bar. 
A bar is a means to drive gas from the outer part of the galaxy towards the centre, as the incoming gas is robbed of its energy, due to shocks. Gas can be funnelled inwards by the bar over  an extended period, so maintaining the star formation in the central region.  Simulations have suggested, however, that a bar can also cause quenching of star formation in the central region by sweeping up the gas within the co-rotational radius within a few rotations, which is then consumed in a vigorous burst of star formation \citep{gavazzi15}. 

Hence, while SMGs and ULIRGs share some similar physical
characteristics, the difference in the extent of their star formation,
and current view of the dust continuum morphologies leaves open the
possibility that  the star formation activity in the two populations
are not driven by the same processes. Indeed, alternative  theories have been proposed  for how cold gas fuels the star formation in high-redshift starburst galaxies  and how the star formation is triggered, through accretion from the cosmic web \citep{bournaud09, dekel09}.  In this scenario these galaxies rapidly accrete gas from the cosmic web and disc instabilities cause clumps to migrate to the nucleus where they form a bulge. These theories predict star-formation rates similar to the gas accretion rate of $\sim100$\,M$_{\odot}$\,yr$^{-1}$ and a resulting morphology of a gas disc twice the size of the nuclear star forming bulge \citep{dekel09}. 

To improve our understanding of the dust continuum structures of
strongly star-forming galaxies at high redshift  and so throw light on
their possible formation and triggering mechanisms, we have analysed
the morphologies of a much larger sample of SMGs to those studied to
date.    In this paper we present the result of this high-resolution
(\res)  continuum study of the \cont\ morphology of \num\ SMGs from
the AS2UDS ALMA survey of sub-millimetre sources in the SCUBA-2
Cosmology Legacy Survey UDS field  \citep{stach19}. This large sample
of uniformly selected SMGs, with integrated continuum signal-to-noise
ratios of $\geq8$, provides a statistically robust constraint on the
sizes and, for the first time, the shapes of this high redshift
population. 
By selecting only the highest resolution
observations
and applying a conservative signal to noise cut, we seek to go beyond
measuring crude sizes for the SMGs and instead derive 
constraints on their profiles and
axial ratios for large statistical samples.   These can then be used
to investigate the physical nature of the dust continuum emission in
these systems.
Our observations resolve the \cont\ dust emission in these sources and
so provide reliable measures of the shape and profile parameters, such
as the effective radius  ($R_{\rm eff}$), \ar\ distribution of the population and typical \sersic\ indices. 
%These distributions indicate that the bulk of the dust continuum emission in a typical SMG arises from a kpc-scale structure with triaxial geometry and an exponential brightness profile -- features which are characteristic of bars.     
%Furthermore, by stacking our sample we recover faint continuum emission from an extended  component which is some two orders of magnitude lower surface brightness, but has an extent nearly an order of magnitude larger than the \cont\ emission detected in the individual sources.  The size of this faint extended component is similar to the typical extent of SMGs as traced by molecular CO gas emission or stellar light seen in   \textit{Hubble Space Telescope} observations.  

An outline of the structure of this paper is as follows:
in \S\ref{sec:obs} we present the observations used for in our analysis.  We analyse these and describe our basic results in \S\ref{sec:res}.  We then interpret and discuss these in \S\ref{sec:dis}, before giving our conclusions in  \S\ref{sec:con}. 
We adopt a standard  concordance, flat $\Lambda$CDM cosmology of H$_0 = 71$\,km\,s$^{-1}$\,Mpc$^{-1}$, $\Omega_\Lambda = 0.73$, and $\Omega_{\rm M} = 0.27$ \citep{spergel07}.

%%%%%%%%%%%%%%%%%%%%%%%%%%%%%%%%%%%%%%%%%%%%%%%%%%
%
%
%
\section{Sample Selection}\label{sec:obs}
Our sample is drawn from an ALMA follow-up study, called AS2UDS 
\citep{stach19}, of the 
sub-millimetre sources discovered in the 
SCUBA-2 Cosmology
Legacy Survey map of the Ultra Deep Survey field \citep[S2CLS][]{geach17}.
Details of the AS2UDS  observations, data
reduction and catalog are given in \cite{stach19}, although we
briefly summarise these here. 
Using ALMA in Cycles 1, 3, 4 and 5 we targeted a complete sample of 716 single-dish
SCUBA-2 850\,$\mu$m sources with observed flux densities of $S_{850}>3.6$\,mJy (corresponding to $>4\sigma$ 
detection significance in the SCUBA-2 map). 
For all the observations, the central
frequency of the receivers was tuned to 344\,GHz and the FWHM of the
ALMA primary beam was 17\farcs3  (encompassing the FWHM of the
SCUBA-2 beam of 14\farcs7).

To reduce the data, we used the Common Astronomy Software Application
\cite[\textsc{casa},][]{mcmullin07} version 4.5.3 using the standard
ALMA calibration scripts.  The data were imaged using the
\textsc{clean}, algorithm in \textsc{casa} with natural weighting
(\textsc{ROBUST} $=2$).  We \textsc{cleaned} the images to the
$1.5\sigma$ level.  Due to configuration differences during these
cycles, the FWHM of the naturally weighted synthesised beam varies
from $0\farcs18$ to $0\farcs35$  (with a small number of repeat observations
obtained at $0\farcs7$   in Cycle 5 to test if flux was being 
resolved out of the higher resolution maps, \citealt{stach19}). Hence to construct the catalogue, all of
the maps were tapered to 0\farcs5 FWHM.  The noise in these tapered
maps varies between $0.09$--$0.34$\,mJy\,beam$^{-1}$
(see \citealt{stach19} for more information about the AS2UDS data
reduction).

The final AS2UDS catalogue contains 706 SMGs that are detected at
$>4.3\sigma$ (2\% false positive rate) with a median flux density
of \flux $\sim3.7$\,mJy \citep{stach19}.  We note that
in the tapered maps, on average we recover the majority of the
single-dish flux for sources with \flux $\gtrsim 3.5$\,mJy \citep{stach19}.  

For this morphological study of the dust emission in SMGs, we
concentrate on the Cycle 3 observations where a subset of 507
SMGs from the AS2UDS survey were detected in maps at a native resolution of
\res\ FWHM.  For relatively high resolution observations similar to these, 
\cite{simpson15b} showed that for a signal-to-noise ratio of
S/N\,$>8$, the uncertainties on the resulting size measurements of sources are
$\lesssim35$\%  \citep{simpson15b}.  We therefore
select all \num\ SMGs which were observed in Cycle 3 (\res\ FWHM) and
are detected with S/N\,$>8$ in the 0\farcs5 tapered maps, and these form the sample for the
remainder of our analysis.    This selection should ensure we can
measure robust sizes and shapes for these sources and that we are sensitive to
a broad range in source sizes.
This sample of \num\ SMGs has a median flux density of \flux $=5.6\pm0.2$\,mJy, roughly $\sim 50$\% brighter than the full sample, with a range in flux density of $2.9$--$11.9$\,mJy, spanning the bulk of the SMG population
which has been studied with ALMA.
This means that though the exposure time of 40 seconds per source is short, we reach similar S/N levels as studies of fainter SMGs but with longer exposure times \citep[e.g.][]{tadaki17}.
%
% Figure 1
%
\begin{figure*}
\centering
\includegraphics[trim=0cm 0cm 0cm 0cm, clip=true,scale=1.0,angle=90]{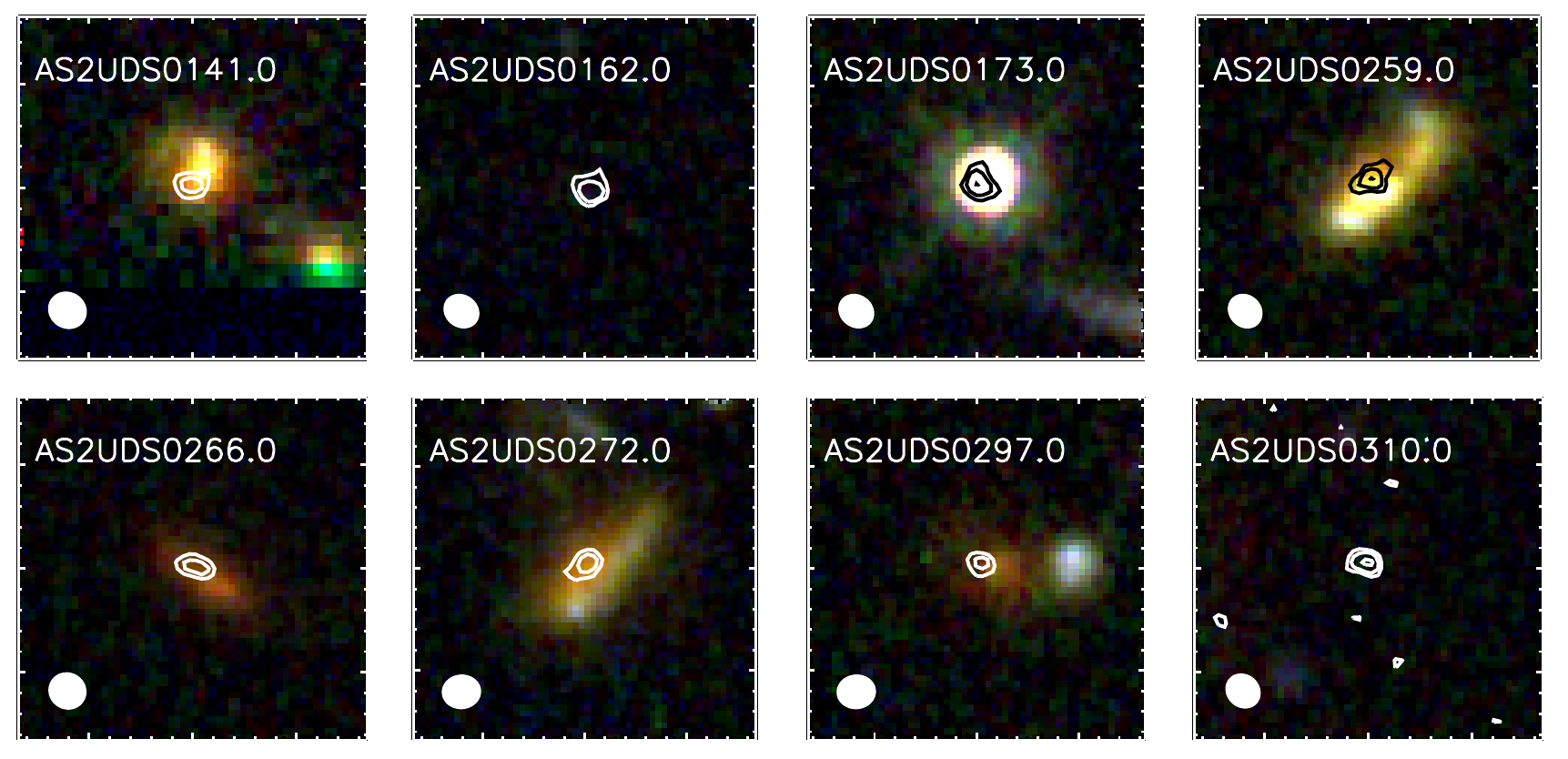}
\caption{{\small \textit{HST} images ($IJH$) of eight examples from the
   \num\ SMGs in our survey, overlaid with \cont\ dust continuum contours at
   $3\sigma$, $5\sigma$, and $9\sigma$, with a median RMS level of $\sim23\,\mu$Jy
The \cont\ dust continuum emission in these eight SMGs 
show some similarities to their stellar emission, in terms of centroid and broad alignment, although 
the \cont\ continuum emission is much more compact
   than the stellar emission. Each thumbnail is
   3\farcs4$\times$3\farcs4, corresponding to $\sim26$\,kpc at the
   median redshift of the sample of $z\sim3$ and the beam size is
   shown in the lower left corner.  }}
\label{fig:maps_hst}
\end{figure*} 

Our ALMA survey was carried out in the $\sim1$\,degree$^2$ UDS
field, part of which was observed with \emph{Hubble Space
 Telescope; (HST)} by the Cosmic Assembly Near-infrared Deep
Extragalactic Legacy Survey (CANDELS).  \citep{grogin11}
This provides F606W, F814W,
F125W, and F160W-band observations for 47 of the AS2UDS SMGs that lie
in the central region of the UDS. We compare the
optical/near-infrared and \cont\ dust continuum morphologies of some of these
galaxies
in Figure~\ref{fig:maps_hst}.  This shows \emph{HST} $IJH$-band colour
thumbnails of eight ALMA SMGs, overlaid with the dust continuum
emission from ALMA.  At the median redshift of our sample
($z\sim\,3$), the observed $IJH$ bands samples the
rest-frame mid-UV to $B$-band, and as Fig.~\ref{fig:maps_hst} shows, the rest-frame
UV/optical morphologies display a range of structures on arcsecond-scales from disc-like to
apparently multi-component mergers, point sources and SMGs which are undetectable in even the
reddest {\it HST} filters.  In contrast, on average, the \cont\ continuum appears much more compact
than the rest-frame UV/optical emission, although generally the emission in the two
wavebands is centred in the same position.
%and aligned along the same major morphological axis. 

\subsection{Multi-wavelength data sets and physical properties of the sample}
Before we assess the dust continuum sizes of our SMG sample, for context we
review the physical properties of our high-resolution sub-sample from
AS2UDS and place them in
context of the parent population of 706 ALMA SMGs in this field.  In
particular, \citep{ugne19} use the extensive
multiwavelength imaging of the UDS to estimate the photometric
redshifts and physical properties of the complete sample, including inferring their stellar masses, star-formation
rates and dust masses.  To achieve this, \citep{ugne19} 
exploit 22-band photometry (or limits) for each
SMG\footnote{The multi-wavelength imaging
 includes photometry  from deep optical $UBVRi'z'$ imaging from
 Subaru and CFHT, near-infrared from  UKIRT ($JHK$) and VISTA ($Y$),
 {\it Spitzer} IRAC\,  3.6--8.0$\mu$m/\,MIPS 24\,$\mu$m mid-infrared imaging,
 deblended far-infrared photometry from \emph{Herschel} PACS
 (100 and 160$\mu$m) and SPIRE (250, 350 and 500\,$\mu$m), ALMA \cont\
 and JVLA 1.4\,GHz}, 
building on the UDS DR11 K-band selected catalogue of Almaini et al. (in preparation), and fit the spectral energy distribution, including deriving the photometric redshift estimates and uncertainties, using the high-redshift version of {\sc magphys} \citep{dacunha15,battisti19}
% and fit the spectral energy distribution using \textsc{magphys} \citep{dacunha08}.

From the analysis of the multi-wavelength SEDs, \citep{ugne19} 
determine that the median redshift of the full sample of
706 ALMA SMGs is $z=2.61\pm 0.08$, with a quartile
range of 1.8--3.4.  The median star-formation rate determined for the
full parent sample is  
235\,M$_{\odot}$\,yr$^{-1}$, the median dust mass is $6.7\times10^{8}$\,M$_{\odot}$ 
and the  stellar mass is $(1.3\pm0.1)\times10^{11}$\,M$_{\odot}$ \citep{ugne19}.
The corresponding values for the  sub-sample of \num\ SMGs in our high-resolution
sample are $z=2.9\pm0.1$, with a quartile range 2.5--3.5, a median
star-formation rate of 380\,M$_{\odot}$\,yr$^{-1}$, a dust
mass of $1.1\times10^9$\,M$_{\odot}$ and stellar mass of $(1.3\pm0.1)\times10^{11}$\,M$_{\odot}$.
As expected, our high-S/N SMG sample, which are roughly 50\% brighter in \flux\ than the 
full sample, also exhibit correspondingly higher dust masses and star-formation rates and due to 
the correlation between observed \flux\ and redshift reported by 
\cite{stach19} means that lie at somewhat higher redshifts than
the full sample.% and have higher star-formation rates.
We will return to a discussion of the  trends of
dust continuum structure with flux and star-formation rate in \S\ref{sec:dis}.

%
% Figure 2
%
\begin{figure*}
\centering
\includegraphics[trim=0cm 0cm 0cm 0cm, clip=true,scale=1.01,angle=90]{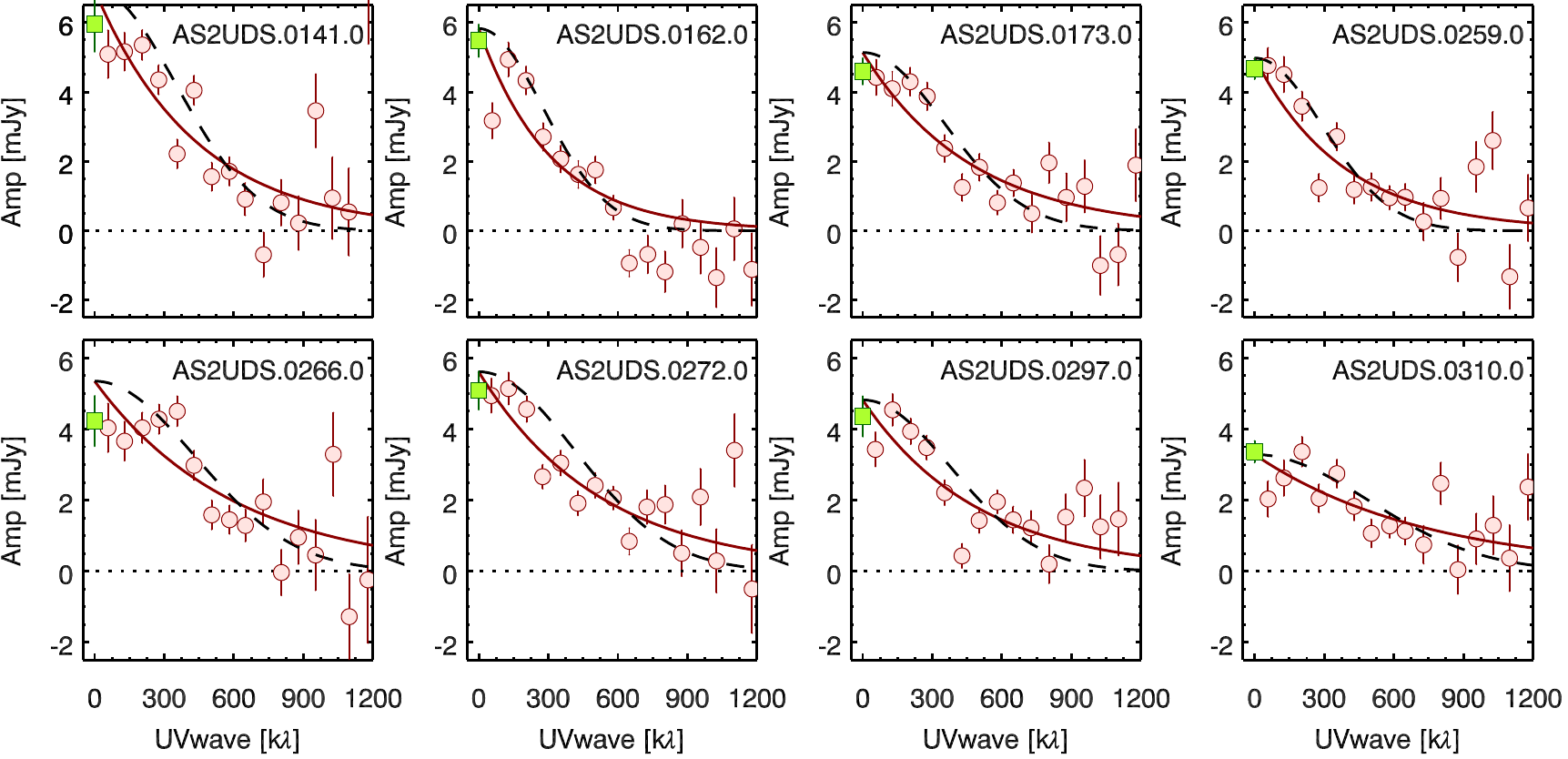}
\caption{{\small Visibility amplitudes (real part) as a function of the
   $uv$-distance  for the eight SMGs shown in
   Fig.~\ref{fig:maps_hst}. The amplitudes are extracted by radially
   averaging the visibilities in 75\,k$\lambda$ bins over the full
   frequency range and the the total flux densities
   recovered in the maps $uv$-tapered to 0\farcs5 resolution are
   plotted as a square. We overlay half-Gaussian fits to the continuum emission
   as a dashed lines, and a \sersic\ fit with $n=1$  by the solid curve.  
   The \cont\ dust continuum of the SMGs
   are all resolved in our observations.  We note that
the Gaussian fits frequently show an apparent compact or unresolved component, 
indicated by a non-zero flux at large  $uv$-distances.  However, this is an
artefact of the fact that the profiles are poorly described by a Gaussian, 
while an $n=1$ \sersic\
better reproduces both the compact and extended emission. 
The \cont\ dust
   continuum sizes of the sources are listed in
   Table~\ref{table:data}.}}
\label{fig:amp_uv}
\end{figure*} 

%
% Figure 3
%
\begin{figure}
\centering
\includegraphics[trim=0cm 0cm 0cm 0.3cm, clip=true,scale=0.85,angle=0]{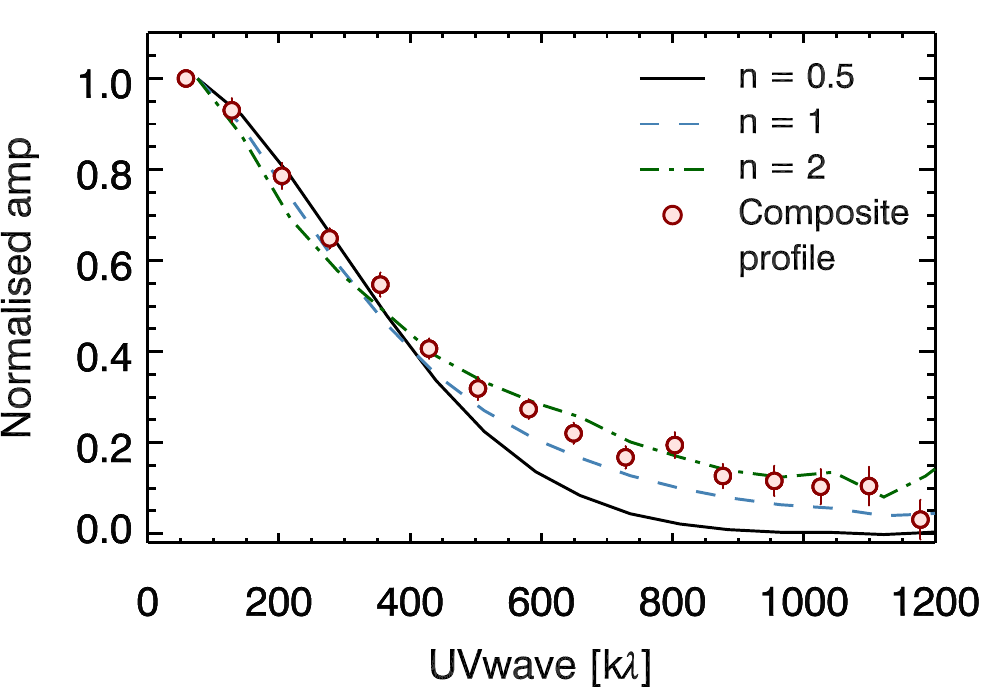}
\caption{{\small Visibility amplitudes as a function of the
   $uv$-distance for three model profiles with $n =
   0.5, 1$ and 2 and radius of 0\farcs15, observed with the same
   array configuration set-up as our observations, compared to the median composite profile of our \num\ SMGs. This illustrate that for a source with an intrinsic Gaussian profile ($n = 0.5$) the amplitude quickly converges to
   zero, while for higher $n$ the amplitude at the largest $uv$ distances
remains non-zero out to at least 1,000 k$\lambda$, mimicking the
signal of a point-source component. This implies that for marginally-resolved galaxies
with profiles steeper than $n>0.5$ the inner part of the galaxy will appear unresolved, incorrectly suggesting the
presence of a point-source component. }}
\label{fig:prof_models}
\end{figure}  

%%%%%%%%%%%%%%%%%%%%%%%%%%%%%%%%%%%%%%%%%%%%%%%%%
%
%
%
\section{Analysis and Results}\label{sec:res}
\noindent

In this section,  we first assess the spatial extent of the 
dust continuum emission in our high-resolution observations
of SMGs  through measurement in both the the $uv$-amplitude plane, and fitting
models to the image plane maps.  We then derive  the \sersic\ profiles
and axis ratios for the continuum emission.

\subsection{Sizes measurements from the visibility plane}

The spatial extent of the \cont\ continuum emission of our
SMGs can be derived by measuring the amplitude as a function of
$uv$-distance.  We apply this approach to each SMG by
first aligning the phase centre of our visibilities with the source
positions from \cite{stach19} using the {\sc casa} task {\sc fixvis} and then radially averaging the amplitudes in 75\,k$\lambda$
bins (the choice of 75\,k$\lambda$ bins is arbitrary, although this
binning minimises the scatter).  In Figure~\ref{fig:amp_uv} we show
the real part of the amplitude as a function of $uv$-distance for the  $uv$-range
out to 1200\,k$\lambda$ for the same eight galaxies shown in
Figure~\ref{fig:maps_hst}. The error bars on the amplitudes are given by the error on the mean in each bin.
In this figure, we also include the total flux measurements from the $0\farcs5$ resolution $uv$-tapered maps.

Figure~\ref{fig:amp_uv} shows that in all cases, the amplitude declines as a function of
increasing $uv$-distance. This is a clear indication that the emission from the source is resolved
in these observations.  Fitting Gaussian light profiles we derive a
median FWHM size of  $0\farcs25 \pm 0\farcs03$.

A subset of our sample were observed at 1.1\,mm with $\sim0\farcs7$ resolution using ALMA by \cite{ikarashi17} 
who studied a sample of millimetre sources selected from the 1.1-mm
AzTEC map of the UDS field.  Unsurprisingly the sources in this bright 1.1-mm sample 
overlap with brighter \cont\ sources in the S2CLS map and as a result 65 of the 69 sources in  \cite{ikarashi17} are also
included in AS2UDS, of
which 30 are in our high-resolution \res\ sub-sample.  We compare the ratio of
the estimated sizes from $uv$-fits to the \cont\ and the lower-resolution (but S/N\,$>10$) 1.1\,mm observations for these 30 sources and derive
a median ratio of FWHM from Gaussian fits of $0.95\pm 0.05$.  This provides strong independent confirmation of the reliability 
of our derived sizes using a completely independent observations, reduction and analysis method.

One noteworthy feature of Figure~\ref{fig:amp_uv} is that it is clear that the amplitude does not
converge to zero at large $uv$-distances in many cases. 
Indeed, in 119 SMGs (out of \num) the amplitude is
non-zero (at $>3\sigma$) at 1200\,k$\lambda$.   Naively this would suggest that
a large fraction ($\sim 80$\%) of sources contain an unresolved component (compared to a Gaussian model)
comprising on average $13\pm 1$\% (or typically \flux $=0.63\pm 0.05$\,mJy) of the emission.

%
% Figure 4
%
\begin{figure*}
\centering
\includegraphics[trim=0.cm 0cm 0.2cm 0.4cm, clip=true,scale=0.86,angle=90]{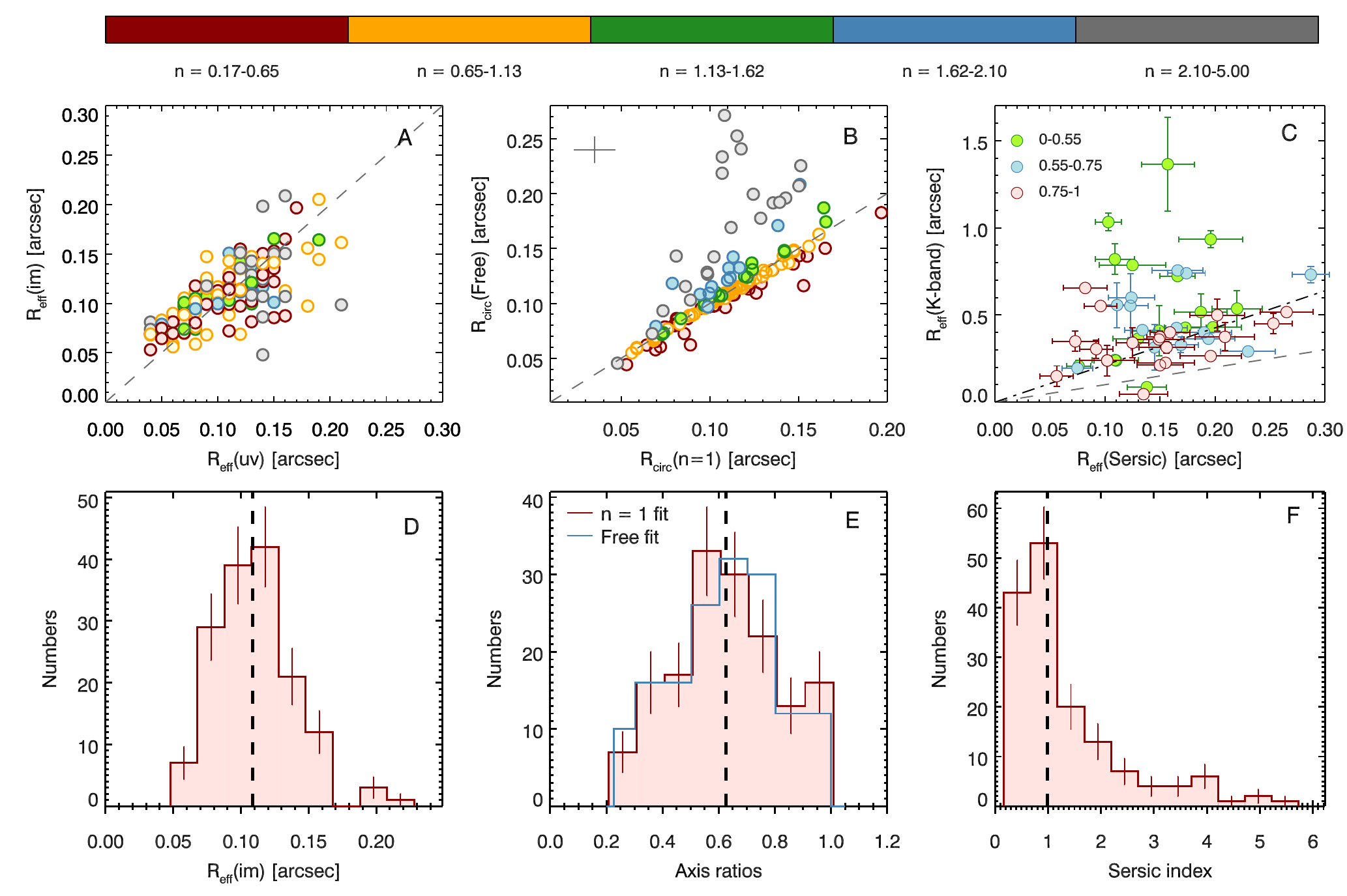}
\caption{{\small \textit{A:} Effective radii (\Reff) values measured in the visibility
    plane (with $n=1$) compared to the circularised \Reff\
    measured using \sersic\ fits with fixed $n=1$ in the image plane.  The \Reff\
    measurements from the two methods show a one-to-one correlation,
    with a scatter partly due to the difference in profile. Profiles
    close to $n\sim 1$ show better agreement than
    profiles with $n\sim 2$--$5$.  \textit{B:} Comparison of the
    circularised \Reff\ ($R_{\text{circ}}$) measurements from
    the $n=1$ and free \sersic\ fits. For $n\lesssim1.6$ the radii
    for the $n=1$ and free \sersic\ fits agree within 1$\sigma$,
    while at $n\gtrsim1.6$ the free \sersic\ fit returns up to
    $\sim2.5$ times higher $R_{\text{circ}}$ than the $n=1$ fit.
    \textit{C:} Effective radii measured in $K$-band imaging as a
    function of \Reff\ we measure with the free
    \sersic\ fit. The points are coloured according to the measured
    \ar s in the $K$-band images. When the sources appear close to circular, \ar s
    $0.75$--1, in the $K$-band image, the measurements show a
 rough correlation (indicated by the dash-dotted curve) with a relative size ratio ($K$ to \cont)  of $2.2\pm0.2$. More elliptical $K$-band measurements show a larger scatter, which (by assuming disc
    morphologies) perhaps indicate that the $K$-band imaging is more affected
    by extinction when viewed edge-on or the geometry of the dust and stellar light is different and hence varies as function of orientation. We return to this point in \S~\ref{sec:model} when we model the structure of the dust in these galaxies and show that the dust morphology does not appear to be disc like.  \textit{D:} Distribution of
    circularised \Reff\ for the $n=1$ \sersic\ fit, with a median \Reff\ of
    $0\farcs10\pm0\farcs0.04$.  \textit{E:} The distribution of the \ar
    s for the $n=1$ fit overlaid with the \ar\ from the free
    \sersic\ fit. The $n=1$ fit distributions has a median of $\sim0.64\pm0.24$, and the free fit distribution has a median of $0.63\pm0.24$. \textit{F:} The \sersic\ index
    distribution of the free fit has a median \sersic\ index of
    $n=1.00\pm0.12$. The majority of the SMGs have $n<1.5$, however, the
    distribution shows a tail out to $n\sim5$.  }}
\label{fig:result}
\end{figure*} 

Large $uv$-distances correspond to small physical scales, and so this
``compact'' emission seen in Figure~\ref{fig:amp_uv} must arise on $\lesssim0\farcs18$ scales.  One option is
that the our observed $uv$--amplitude profile comprises  a luminous, extended
(Gaussian-like) starburst with a $\sim13$\% contribution from a central point-source
\citep{tadaki17}.  However, it is also possible that the ``compact''
emission instead arises from a light profile which is more centrally
concentrated than a Gaussian (which is generally a poor description
of the light or mass profiles of resolved galaxies).  To investigate how
different light profiles should appear in the $uv$-amplitude plane, in
Figure~\ref{fig:prof_models} we show the simulated $uv$-amplitude-distance as a
function of \sersic\ index ($n$) with $n=0.5$, 1 and 2 and the median composite profile of the \num\ SMGs in our sample.
These profiles were created using the {\sc casa} simulation tool with the
same configuration as our observations (and hence the same synthesised
beam FWHM).  For brightness profiles with \sersic\ index $n>0.5$ the $uv$-amplitude profile does not converge
to zero by a $uv$-distance of $\sim1200$\,k$\lambda$ -- a consequence of the steep central
light profile which gives rise to apparently compact emission.  
Hence, fitting a Gaussian model to a marginally resolved source with an intrinsic \sersic\ $n=1$ profile you would  conclude that a second compact emission component was implied by the non-zero amplitude at large $uv$-distance.
To illustrate the difference in Fig.~\ref{fig:amp_uv} we overlay the
best-fit models with $n=0.5$ and $n=1$ to the eight galaxies
shown. The $\chi^2$ distributions of the individual $n=0.5$ and $n=1$ fits show a moderate preference ($2.5\sigma$) towards the $n=1$ fits. This is supported by the composite profile, which is best fit with a $n=1$ profile. The $n=1$ fits results in $\chi^2\sim1.5$, compared to $\chi^2\sim2.5$ for a $n=0.5$ fit.
%To highlight the difference, in Fig.~\ref{fig:amp_uv} we overlay the
%best-fit models with $n=0.5$ and $n=1$ to the eight galaxies shown.  
%In all but one case (AS2UDS\,0266.0) the $n=1$ fit provides a significantly improved fit.  
As we show in \S~\ref{sec:size_im} fitting a \sersic\ model to each of the SMGs in our sample suggests
a median \sersic\ profile of $n=1.00\pm0.12$.
This indicates that the majority (77\%) of the SMGs in our sample have
\cont\ continuum profiles that are consistent with
\sersic\ with $n\sim 1$ (rather than Gaussian, $n\sim 0.5$) light
profiles. 
We therefore conclude that \sersic\ models provide an appropriate 
description of the \cont\ brightness profiles of the SMGs in our
sample. 
Hence to measure the spatial extent of the dust continuum emission, we adopt
an $n=1$ \sersic\ and allow the effective radius as a free
parameter and fit the $uv$-amplitude profile for each SMG.  We derive
a median effective radius for the  \num\ SMGs in our
high-resolution/high-S/N ratio sample of $0\farcs10\pm0\farcs04$.

The typical effective radius we measure is comparable to estimates
from previous studies.  For example, \cite{hodge16} measure the
spatial extent of the \cont\ dust continuum in 16 SMGs from ALMA
observations with $\sim0\farcs16$ resolution and a sample with flux density range of \flux$=3.4$--$9.0$\,mJy.  
They derive a median effective radius of $0\farcs15\pm0\farcs03$, which is similar to our sample.  
\cite{simpson15a} also measure the \cont\ sizes of 30 from our ALMA Cycle 1
observations of the brightest SMGs in
the AS2UDS pilot (there  is no overlap in
sources to the \num\ SMGs analysed here).
Using that $\sim0\farcs3$ resolution data,
\cite{simpson15a} derive a median effective radius of $0\farcs13\pm0\farcs02$.  
While, as noted earlier, \cite{ikarashi17} analysed ALMA 1.1-mm
observations of 65 SMGs from AS2UDS and derive an effective
radius of  $0\farcs13\pm0\farcs06$ using their lower resolution,
$0\farcs7$ data.   Thus it appears that the extent of the dust continuum emission we
measure for our sample  is comparable to that estimated from
earlier smaller-scale  studies.

\subsection{Sizes and shape measurements from the image plane}\label{sec:size_im}
The azimuthally averaged amplitude measured from the visibility plane is well suited for deriving a
characteristic radius for the emission from a galaxy, but provides a
circularised average.  Information about the \sersic\ index and the
axis ratio ($b/a$) of the emission can also be derived, in a
computationally more tractable manner, from the image plane.  

To measure the sizes from the image-plane maps of each SMG we fit a two
dimensional \sersic\ surface brightness profile, allowing the
effective radius,
axial ratio and \sersic\ index to vary.  We account for the beam
by convolving each intrinsic model with the
synthesised beam with a semi-major and semi-minor axes and position
angle given by the beam parameters for each map.  The fit returns
measurements of the peak flux, the central position, the semi-major
axes, the \ar\ (i.e. the ratio of the minor to major axes, $b/a$), the
position angle, and for the case of free \sersic\ fit, the
\sersic\ index ($n$) as well as uncertainties on all these parameters. 

Before we present the results of the fitting, we first test the
reliability of the deconvolved measurements, and calibrate their
uncertainties.  To do this, we generate a set of 1,000 simulated
galaxies using \textsc{casa} which have a flux distribution similar to
our sample.  These simulated galaxies have semi-major axis between
0\farcs11 and 0\farcs24, and random inclination angles and \sersic\ index.  We use \textsc{casa} to
simulate the observations of these galaxies with the same exposure
time as our data, and hence these simulated maps have similar noise
properties as our sample.  We then fit these simulated galaxies with
our code and return their best-fit parameters.  
On average we recover the effective radii to within 20\% and the \ar s\ are recovered within 12\% of the value of the input parameters.
%$\Delta$\Reff/$R_{\text{eff,in}}=0.20\pm0.01$. 
%The \ar s are recovered with $\Delta(b/a)/(b/a)_{\text{in}}=0.12\pm0.08$.
The \sersic\ index is the most difficult parameter to
fit and recover at the signal to noise of our typical sources, with  a typical error of 22\%. %$\Delta n/n= 0.22\pm 0.01$.
To investigate the potential bias due to noise when measuring the shapes of round sources (which may cause an observed decline in the apparent numbers of round sources), we also test the code on Gaussian profiles with \ar\ $b/a= 1$ and find that the procedure with free fit returns the \sersic\ index within $25$\% and the \ar\ within $14$\%.
The \ar\ distribution peaks for $b/a=0.86$ and has a standard deviation of $0.12$ for a sample matched in signal-to-noise to our observations. We therefore conclude that the strong peak in \ar\ at $b/a\sim0.65$ is our observed distribution if not a result of this bias.

The measurement of the \sersic\ can also be influenced by data
sampling \citep[e.g.][]{robotham16}, and so next we investigate the
influence of the reconstructed map pixel sampling by testing the same procedure with
different pixel scales.  We create 1,000 model maps of SMGs at
the same pixel scale as our data ($0\farcs03$\,pixel$^{-1}$), and also
at three times smaller sampling ($0\farcs01$\,pixel$^{-1}$).  We simulate
observations of these maps with \textsc{casa} and then fit these these
maps and infer their properties.  This test shows that finer oversampling
of the synthesised beam does not return more accurate or precise \sersic\ indices
$\Delta n/n= 0.34\pm 0.05$.  For low signal-to-noise profiles with input
\sersic\ indices $n\gtrsim1.25$ the fitting-procedure on the over-sampled maps return
systematically lower values of $n$, but with
increasingly larger uncertainties.

To derive measurements of the effective radii of the dust continuum in
our SMGs, we now perform two sets of fits.  First, we perform a fit
with \sersic\ $n$ as a free parameter.  In these fits, we derive a
median $n=1.00\pm0.12$ and \Reff$=0\farcs11\pm 0\farcs01$.
Since the \sersic\ index is the least certain parameter, we
then fix the \sersic\ index to $n=1.0$ and refit each SMG.  For this
$n=1.0$ fit, the median effective radius for the sample is
\Reff$=0\farcs10\pm 0\farcs04$ (see Fig.~\ref{fig:result}).

In Fig.~\ref{fig:result} we compare the effective radius for the SMGs
derived from the $uv$-fitting with that derived in the image plane.
We first compare the effective radii derived from the fixed $n=1$ \sersic\ fit
in both cases, deriving a median ratio of the image-plane to
$uv$-plane of \Reff($uv$)/\Reff(im)$=1.10\pm 0.01$.  Although the two measurements are
correlated, it is also clear from Fig.~\ref{fig:result} that when fitting a
\sersic\ model in the image plane, for larger effective radii, the $uv$-derived effective
radius is $\sim30$\% larger than that derived in the image-plane,
but with no strong trend with \sersic\ index.

In Fig.~\ref{fig:result} we compare the effective radii of the dust continuum in
the image-plane for the free and fixed \sersic\ models.  The median
ratio in effective radii of \Reff(free)/\Reff$(n=1)= 1.07\pm0.01$.  The scatter in this relation
can be attributed to those SMGs with profile with a higher \sersic\ (as
indicated by the colour scaling of the points, which show that for
higher \sersic\ index, the free fit returns larger sizes).  We also
show the \sersic\ index distribution derived from the free fit, which
has a median of $n=1.00\pm0.12$.  This shows that $\sim70$\% of
the SMGs in our sample have \sersic\ indices $n=0.7$--$2$ (as also
suggested by the $uv$-amplitude profiles).

Using the image-plane fits, we extract the distribution of axis ratios
from the best-fit models from both the free and fixed \sersic\ model
fits, and show these in Figure~\ref{fig:result}.  Both distributions
are strongly peaked, with a median of $b/a=0.63\pm0.02$ for the fixed $n$ fit and
$b/a=0.64\pm0.02$ for the free $n$ fits, and both distributions have
the same standard deviations of $\sigma_{\text{b/a}}=0.19$.    We
confirm that there is no correlation between measured $b/a$ and
effective
radius. 

\cite{hodge16} present axial ratios for their 16 ALESS SMGs  observed
at \cont\ with
$0\farcs16$ resolution, which span a narrow range in $b/a=0.3$--$0.7$ and
 have a median of $b/a=0.55\pm 0.06$ and $\sigma_{\text{b/a}}=0.13$.
Their distribution is thus consistent with that from our larger sample,
 and both display a relative lack of ``round'' sources (compared to
 the naive expectation of randomly orientated circular, thin discs, \citealt{ryden04}).
We will return to a discussion of this distribution in
\S\ref{sec:dis}.

In summary, both the \cont\ dust continuum in the $uv$-amplitude
profiles and image-plane maps suggest that the majority of the
SMGs in our sample are best fit by $n\sim 1$ \sersic\ models.  Fitting
\sersic\ models to the $uv$ and image plane returns consistent
results, and suggests a median effective radii of $0\farcs10\pm0\farcs04$
($\sim 0.8\pm 0.3$\,kpc at the median redshift of our sample). 
Since the \sersic\ fit in the image-plane maps also allows us to easily
investigate the shape parameters for the SMGs, for the remaining
analysis we will use this method, but will adopt the minimum
uncertainties on any measurement of effective radius from the scatter
determine from the correlation(s) in Fig.~\ref{fig:result}.

\subsection{Comparison of the dust and rest-frame optical emission}
In Fig.~\ref{fig:maps_hst} we show the \emph{HST} images of eight
example SMGs in our sample which are also observed as part of the
CANDELS survey \citep{grogin11}.  It is clear from this figure that the rest-frame
optical emission is much more extended than the \cont\ emission
\citep[see also][]{simpson15a, chen15, lang19}.
To quantify this further, we perform a \sersic\ fit to the $K$-band
images of all of the SMGs in our sample using GALAPAGOS (\citealt[][Maltby et
al.\ in prep]{almaini17}) which takes the PSF into account, and we compare the
  effective radii measured in the UDS $K$-band compared to that
  measured from \cont\ in Fig.~\ref{fig:result}.  This shows that
  for galaxies with $K$-band \ar\ of $b/a=0.75$--1
  (i.e.\ close to circular) the $K$-band effective radii are a factor
  of $2.2\pm0.2$ larger than the effective radii measured at \cont, although with considerable scatter (especially for those with large apparent axial ratios). 
  This implies that the stellar light distribution is typically twice that of dust \citep[see also][]{lang19}.  We will discuss the extended emission in the stars and dust further in \S\ref{sec:dis}. 
  We note, however, that optical depth effects need to be considered, in cases of more detailed comparisons of sizes measured at different wavelengths, for example between short wavelength dust emission, and \CII\ or low or high-$J$ CO emission.

\subsection{Size and shape evolution}

%
% Figure 5
%
\begin{figure*}
\centering
\includegraphics[trim=0cm 0cm 0cm 0.8cm, clip=true,scale=0.86,angle=0]{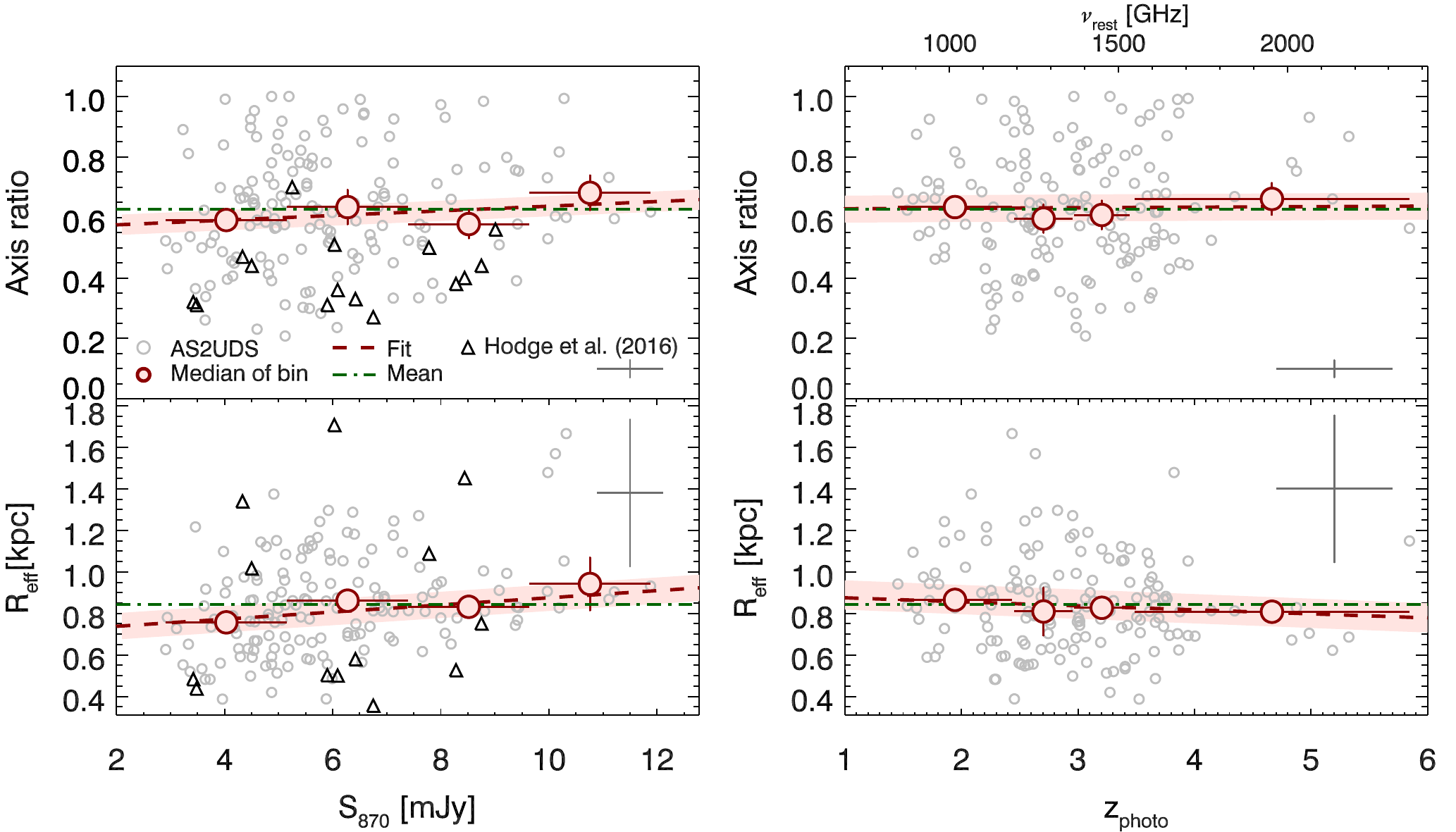}
\caption{{\small \textit{Left:} Measured \ar s (top) and \Reff\ (bottom) from the $n=1$ fit as functions of the \cont\ continuum flux density. The large points are the medians in bins of equal numbers of sources.
We see weak positive correlations between both the \ar\ and \Reff\ with \cont\
continuum flux density (dashed curves).
\textit{Right:} Measured \ar s (top) as function of the photometric
redshift (\photz; \citealt{ugne19}) shows a very weak positive
correlation, while the \Reff\ (bottom) decreases marginally as a
function of the photometric redshift.  
These data are observed \cont\ meaning that the rest-frame wavelengths
vary across the redshift range $z\sim1$--6, and hence the effective
optical depth is expected to vary at the observed wavelength. 
The influence of these optical depth effects, as well as potential
evolution in the structure and physical properties of the SMGs (e.g.\ dust mass and
far-infrared luminosity) mean it is hard to draw strong conclusions from
this plot.  Nevertheless, we suggest that the lack of any strong trends
in effective radius with redshift most likely indicates that there is no
{\it strong} evolution in the size of the SMGs with redshift.
}}
\label{fig:relations}
\end{figure*} 

Previous morphological studies (with sufficiently high signal-to-noise
detections) of SMGs at sub-/millimetre wavelengths have been limited
either by moderate resolutions \citep{simpson15b, ikarashi17} or
modest sized samples \citep{hodge16, gullberg18, hodge19}. 
Our sample of \num\ SMGs detected at S/N\,$>8$ in \res\ resolution maps, allow for a statistical study of the morphology, sizes and \ar s for a wide range in both \cont\ flux density (\flux $\sim 2.7$--$11.5$\,mJy) and redshift ($z\sim1$--$6$). 

Our sample has a median redshift of $z=2.9\pm 0.1$ and a quartile range
in redshift of 2.5--3.5, so the corresponding rest-frame wavelength
($\lambda_{\rm rest}\geq$\,150–-300\,$\mu$m) of the dust emission from
the galaxies should be generally optically thin \citep[e.g.][]{simpson17}.
However, at the highest redshift, the dust emission may be optically
thick (e.g.\ at $z\gg3.5$--$4$, the rest-frame wavelength is
$\lambda<200\,\mu$m),  this would then produce a small bias such that
galaxies whose orientation provides a larger apparent sky area (i.e.\
face-on for disc-like geometries) would have brighter \flux .
There is a possible hint of this in Figure~\ref{fig:relations}
where we plot the \ar s as a function of both \cont\ flux density and
redshift \citep{ugne19}, and find weak positive correlations. 

In the plot of effective radius (\Reff) versus \cont\ flux
density, we identify a weak positive correlation.  But this trend
is marginal and so  we conclude that the
brighter SMGs are more luminous primarily due to their higher dust surface densities,
rather than their larger sizes.   The apparent trend suggests a doubling
of dust mass surface density between SMGs with \flux $\sim 4$--$11$, which
may imply a similar increase in gas density and a corresponding 
rise in mid-plane pressure
in these systems, which are already thought to be extremely high \citep{swinbank11, swinbank15}.  

More interestingly, we see a small variation of the effective radius of $\sim10$\% (corresponding to $\sim2.5\times10^8$\,M$_{\odot}$\,kpc$^2$), by a weak decline with redshift \citep{ugne19}.  We expect this behaviour to reflect both evolution in the physical size of the sources and also the influence of dust optical depth, dust temperature and source structure.  For simple source geometries, the apparent sizes of sources are expected to 
decline with increasing $L_{\rm FIR}/M_{\rm d}$ 
\citep{scoville13}.  Using the estimates of $L_{\rm FIR}$ and $M_{\rm d}$ for our sample
from \citep{ugne19}, we expect an increase in median $L_{\rm FIR}/M_{\rm d}$ of only $\sim20$\% in the sample across $z\sim2$--4 \citep{scoville13}, which would
correspond to a $\sim20$\% decline in apparent size at a fixed
rest-frame wavelength.    However, this will be countered by the effect
of shifting to shorter (and hence optically thicker) rest-frame wavelengths as we observe higher
redshift sources.  Thus we expect the drop in observed \cont\ effective radius with
redshift to be less than $\leq 20$\%, which would be consistent with
the weak decline seen in Fig.~\ref{fig:relations}.

Overall, we conclude there are a number of potentially competing effects which could influence
the variation in apparent size of the SMGs with redshift, but none of these effects is strong and
hence the absence of significant evolution in the effective radius with redshift
most likely indicates that the intrinsic physical sizes of the SMGs do not 
evolve strongly with redshift.   This would be in contrast with 
studies in the UV and optical of a variety of galaxy populations which
have reduction of a factor of several in typical size with redshift for both quiescent
galaxies and star-forming galaxies across $z\sim 0$--$4$ \citep[e.g.][]{vanderwel14a, shibuya15, kubo18}.

%%%%%%%%%%%%%%%%%%%%%%%%%%%%%%%%%%%%%%%%%%%%%%%%%%
\subsection{Modelling the \ar\ distribution}\label{sec:model}

Our results above, as well as recent studies \citep[e.g.][]{simpson15b, hodge16}, have shown that the \cont\ dust emission in SMGs follows an exponential 
surface brightness profile suggestive of a disc-like geometry. 
For a sample of circular exponential discs viewed at random viewing angles, the \ar\ distribution should be constant at high axial ratios, with a decline towards $b/a=0$, the strength of
which depends upon the relative thickness of the disc.
However, as shown in Fig.~\ref{fig:result} for our SMGs, the apparent
axial ratio distribution is highly peaked, with  a large fraction of the sample ($62\pm4$\%) having \ar\ in the range $b/a=0.5$--$0.8$, and proportionally fewer 
at $b/a<0.5$ and $b/a>0.8$, where the latter account for $15\pm8$\% of the sample.  This
``deficit'' in the number of  sources with near-circular shapes
suggests that either the assumption of SMGs being circular discs with
a uniform viewing angle distribution is incorrect, or that
signal-to-noise effects cause us underestimate the numbers of high
$b/a$ sources. 

We first confirm that the behaviour we see is not due to
signal-to-noise effects.   
To ensure that our profiles are robust, we set a lower limit for the signal to noise of S/N $>8$, since the fractional uncertainties on the measured sizes increases to over $\sim35$\% at lower signal-to-noise ratios \citep{simpson15b}. 
The signal-to-noise ratio of our sample ranges from 8 to 29 with a median of 12. 
Dividing the sample in half at S/N $\simeq12$ yields two distributions which both show a ``deficit'' of \ar s at $>0.8$, suggesting that the signal-to-noise  is not the cause of the observed ``deficit'' of high \ar\ sources.
Hence we now explore three geometrical models for the sources to attempt to reproduce the 
observed \ar\ distribution (including the influence of potential selection effects):  optically thick discs, optically thin
discs and a model with a triaxial geometry for the sources.

%
% Figure 6
%
\begin{figure*}
\centering
\includegraphics[trim=0cm 0cm 0cm 0cm, clip=true,scale=0.85,angle=90]{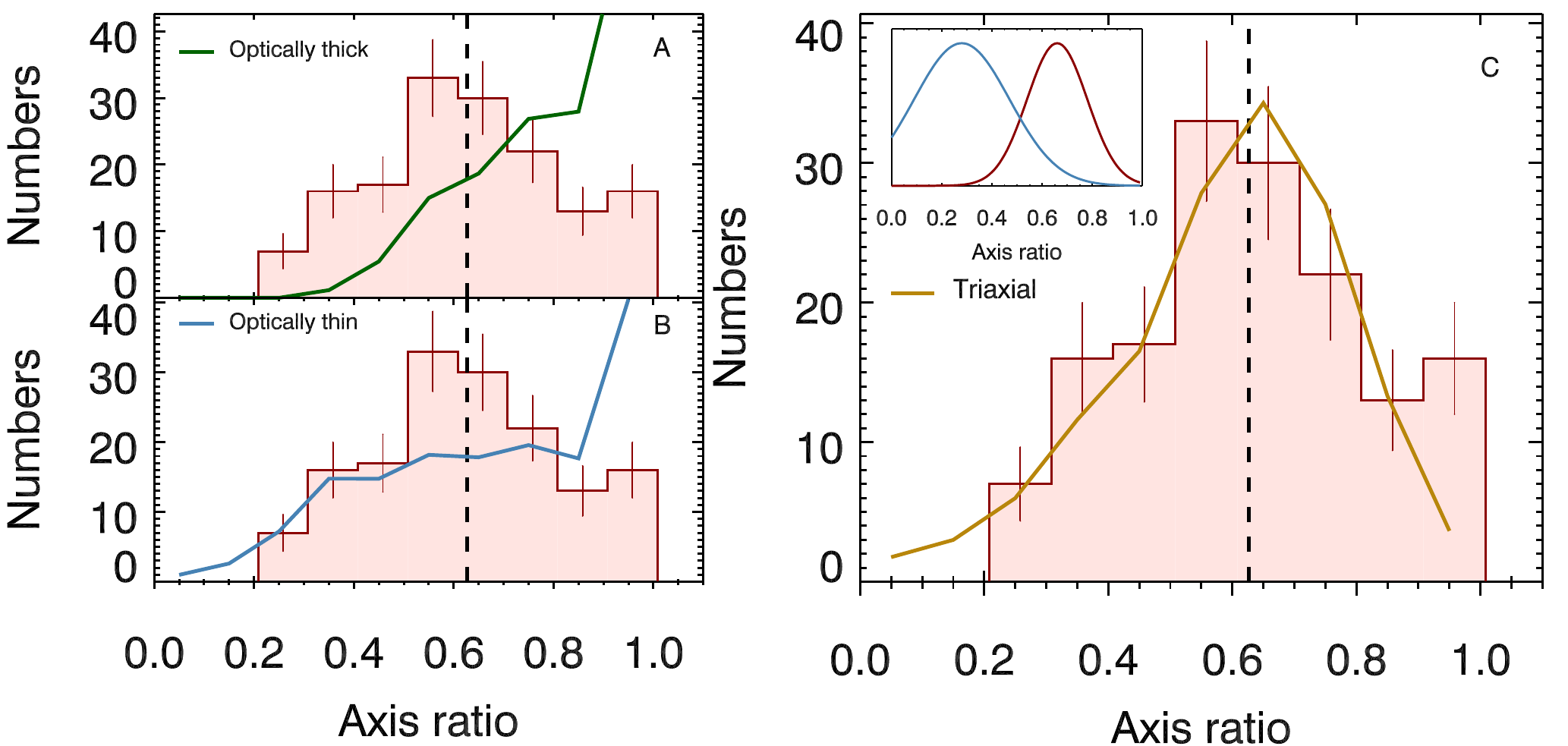}
\caption{{\small Distribution of the \ar s for $n=1$ compared with the expected axis ratios from $a$) optically thick, $b$) optically thin disc models and $c$) from a triaxial model.
For the \ar\ distribution of the optically thick and thin models to
have a similar ``deficit'' of high axis ratios (face-on discs), we
would have to assume an unphysical model where we
are preferentially viewing the sources at a particular viewing angle e ($\sim40^{\circ}$). 
In contrast the triaxial model (with a uniform viewing angle
distribution) can broadly reproduce the observed \ar\ distribution for our sample of SMGs. 
\textit{Right insert:} In the best-fit triaxial model the intrinsic $b/a$ and $c/a$ distributions that result in the observed axis ratio distributions are 
Gaussian distributions with a peak at $0.68\pm0.02$ and a width of $0.12\pm0.06$ for $b/a$ and a peak at $0.28\pm0.01$ with a width of $0.19\pm0.01$ for $c/a$.}}
\label{fig:models}
\end{figure*} 
%ZZZ WHY DO THE MODEL HISTOGRAMS INCREASE RAPIDLY AT B/A>0.85?!?!?!

\subsubsection{Geometrical models}
To assess the possible influence of selection effects on the \ar\ distribution we
compare our observed \ar\ distribution with those predicted for
sources which are modelled as optically thick or thin circular discs. We generate a
simulated sample of circular discs, where the apparent \ar\ is only dependent
on the (random) viewing angle. We follow the example of \cite{ryden04}, where the apparent \ar\ $q$ is given by
\begin{align}\label{eq:q}
q=\left[\frac{A+C-\sqrt{(A-C)^2+B}}{A+C+\sqrt{(A-C)^2+B}}\right]^{1/2}, 
\end{align}
where, $A$, $B$ and $C$ are given by
\begin{align}
A &= [1-\epsilon(2-\epsilon)\sin^2\phi]\cos^2\theta+\gamma^2\sin^2\theta, \nonumber \\
B &= 4\epsilon^2(2-\epsilon)^2\cos^2\theta\sin^2\phi\cos^2\phi,\nonumber\\
C &= 1-\epsilon(2-\epsilon)\cos^2\phi.\nonumber
\end{align}
Here, $\epsilon$ is the ellipticity of the source $\epsilon=1-b/a$ (where $a$ and $b$ are the intrinsic major and minor axes), and $\gamma$ is the ratio between the third axis ($c$) and the major axis; $\gamma=c/a$. The two angles $\theta$ and $\phi$ are the two viewing angles. 
Only $\theta$ has an influence on the apparent \ar\ in the case of a circular disc, where $a$ and $b$ are equal. 

The resulting distribution of apparent \ar s for circular discs is therefore dependent on the distribution of the viewing angle $\theta$, the $c/a$ ratio, and the flux distribution. We assume a flux distribution similar to our sample and calculate the apparent \ar s for the possible combinations of the different parameters, adopting  a uniform distribution for the viewing angle and for $c/a$.

\noindent{\it Model I.\ Optically thick disc} ---
In the case of an optically thick disc, at an observed wavelength of \cont, the fraction of the emitted emission that is detected by the observer is given by the visible fraction of the surface area (i.e.\ the apparent \ar). 

We attempt to fit the distribution of $b/a$ with Eq.~\ref{eq:q} for this optically thick model, but in all cases the best-fit is poor (see Fig.~\ref{fig:models}).
The best-fit optically thick disc models with a uniform viewing angle distribution has a Gaussian-shaped $c/a$ distribution, peaking for $c/a=0.06$ with standard deviation of 0.33. 
However, the best-fit  is still poor, and a Kolmogorov-Smirnov (KS) test shows a negligible (0\%) chance for these two distributions to be drawn from the same parent sample. 
While it would be possible to bias the viewing angle distribution and
find a better fit, such non-uniform viewing angle distributions are
unphysical in the absence of an identifiable selection bias as a
cause, and so we discard this option.\footnote{We note that our sample
  is selected from low resolution single-dish observations and so is
  not expected to 
  suffer from surface brightness selection biases.}

\noindent{\it Model II: Optically thin disc} ---
For the optically thin case the emission only depends on the assigned
source brightness. 
As for the optically thick case, we are unable to find acceptable fit
to the observations for the optical thin model with a uniform viewing angle distribution. 
The closest model for the optically thin case, with a uniform viewing angle distribution, has a rise in \ar\ distribution beginning at lower \ar s, and has a flatter and slower rise than in the optically thick case. 
The best-fit model with a uniform viewing angle distribution has a
$c/a$ distribution peaking at 0.09, with a standard deviation of
0.28. Again a KS test returns a negligible chance (0\%) of the two distributions being drawn from the same parent sample.
The only way to match the ``deficit'' at high \ar\ would then be to include a biased  viewing angle distribution, which as noted earlier is not physically plausible.

\noindent{\it Model III: Triaxial structure} ---  \label{sec:tria}
The shape of our observed \ar\ distribution differs from those seen
for late-type spiral galaxies  \citep{ryden04}, which lack the strong
peak at $b/a\sim 0.6$ and the associated
deficit at high \ar s which we observe.   We have quantitatively
confirmed this above and conclude that 
neither of the circular disc models is able to adequately fit the data
without invoking assumptions of unphysical viewing angle biases.  Thus we now explore a triaxial model, where $a>b>c$. 
In this model we determine the best fit $b/a$ and $c/a$ \ar s distribution to fit the observed \ar\ distribution for an optically thin case. 
For the triaxial case both viewing angles ($\theta$ and $\phi$) have an influence on the apparent \ar\ $q$. 
We assume uniform angle distributions for both $\theta$ and $\phi$, and model the apparent \ar\ distribution for a range of model and width parameters for the $b/a$ and $c/a$ distributions. 
We assume that the $b/a$ and $c/a$ distributions follow Gaussian distributions given by
\begin{align}
f(q_{\text{int}})=\exp\left(-\frac{(q_{\text{int}}-q_0)^2}{2\sigma^2}\right),
\end{align}
where $q_{\text{int}}$ represents the intrinsic $b/a$ and $c/a$ \ar s, $q_0$ is the mode value, and $\sigma$ is the width of the distribution. 
We calculate \ar s for two different ranges of mode and width parameters for $b/a$ and $c/a$ (see Table \ref{table:triax}).

%
% Figure 7
%
\begin{figure*}
\includegraphics[trim=0cm 0cm 0cm 0cm,
 clip=true,scale=0.855,angle=0]{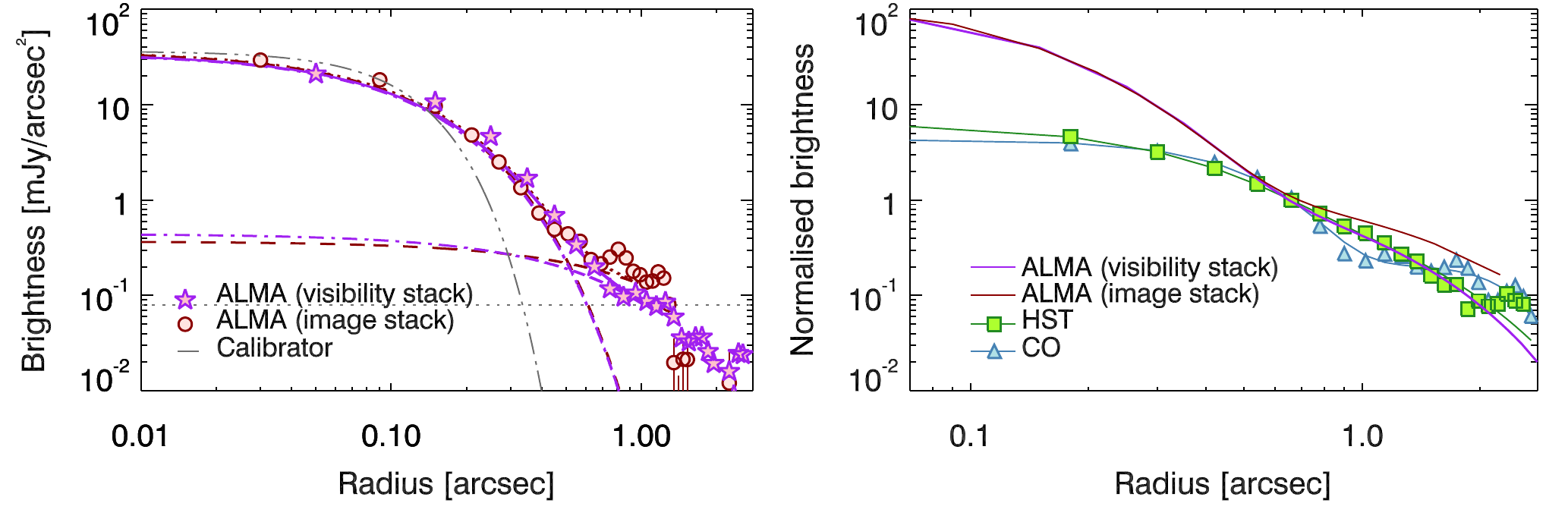}
\caption{{\small \textit{Left}: The \cont\ continuum light profile obtained by
   stacking the \cont\ dust maps in the visibility and image
   plane. The profile shows that most of the emission is compact, but that $13\pm1$\% of the emission is extended
   on larger scales ($\geq 0\farcs5$ or $\geq 10$\,kpc), and only detected in
   our maps by stacking. The RMS level of the image stack is illustrated by the dotted line.
   \textit{Right}: The \cont\ continuum light profile obtained by
   stacking the \cont\ dust maps in the visibility and image planes, normalised and compared to the stacked
   stellar and molecular gas profile seen in \textit{HST} and ALMA
   imaging \citep{rivera18}.  While the \cont\ dust continuum
   emission has an excess in the nuclear region compared to the
   stellar and molecular gas profiles, the extended component follows
   that of the stellar and molecular gas, suggesting that they trace
   the same structural component in these systems.  }}
\label{fig:stack}
\end{figure*}

%
% Table 1
%
\begin{table}
\centering
\begin{tabular}{l  c c }
\hline\hline       
Parameter         & Full range   & Best fit \\%Binsize$_1$, Binsize$_2$ \\
$q_{b/a}$          & 0.1--1     &  $0.66\pm0.02$ \\%$0.1,0.02$  \\
$\sigma_{b/a}$ & 0.3--3      &  $0.12\pm0.06$ \\%$0.3,0.06$\\
$q_{c/a}$          & 0.03--0.3 & $0.28\pm0.01$ \\%$0.03,0.006$\\
$\sigma_{c/a}$ & 0.05--0.5 & $0.19\pm0.01$ \\%$0.05,0.01$\\
\hline
\end{tabular}
\caption{{\small Parameter ranges that we apply to explore the triaxial model, and the best-fit values for the model.
}}
\label{table:triax}
\end{table}

As the distributions are expected to be continuous and smooth, we
perform our search for the best-fit model by running through the
parameter space twice; first by using a large bin size to cover the
full parameter range and select the combination of parameters that
provide the closest match to the observed \ar\ distribution; this
best-fit parameter combination is then used as the centre of the second run which uses a finer search grid. 
The parameters for the $b/a$ and $c/a$ distributions that result in an apparent \ar\ distribution best fit to our observed distribution (and see Fig.~\ref{fig:models}) are given in Table~\ref{table:triax} (see also Fig.~\ref{fig:models}).
A KS test shows that there is a 40\% chance that the triaxial \ar\ distribution and our observed \ar\ distribution originate from the same parent sample.
This is further supported by 
the Akaike Information Criterion (AIC), which takes into account the number of fitted parameters, and for the triaxial case is 10.8. 
The AIC values for the optical thick and thin cases with uniform viewing angle (discussed
in \S 4.4) distributions are 33.8 and 15.2.
The model resulting in the lowest AIC values yields the best-fit model, which is therefore the triaxial model.
We conclude that the ``deficit'' in high \ar s is most likely due to
intrinsic triaxial morphologies, rather than the dust continuum
emission  of SMGs resembling randomly orientated circular disc galaxies. 

\subsection{Stacked emission profiles}

Near-infrared \emph{HST} imaging of ALMA SMGs shows that the rest-frame
UV/optical emission is extended on $\sim8$--$10$\,kpc scales
(e.g.\ Fig.~1; see also \cite{chen16, lang19}).  
In comparison, as we showed above, the \cont\ dust
continuum is much more compact, with an effective radius of just
$0\farcs10\pm0\farcs04$ or $\sim1$\,kpc. But is a more extended, lower
surface brightness, emission component also present?
  
\cite{stach19} showed that our \res\ ALMA resolution observations of the \cont\
emission from the SMGs  recover $\sim95$\%     
of the single-dish flux detected with
SCUBA-2.  This may indicate that a small fraction of
the flux is resolved on larger scales, but we do not have the
sensitivity to detect this extended emission on a case-by-case basis.
Instead to search for this emission we can stack the \cont\ continuum maps of
the SMGs.   As with our size measurements, we perform this stacking 
in both the visibility and image planes to assess the reliability of
our results.

First, we stack the SMGs in the visibility plane, which has the
advantage of circumventing issues arising from inhomogeneous
beam-sizes of the individual maps.  
We shift the phase centre of the ALMA primary beam to correspond to 
the position of the SMG (all of our targets
are the sole detected SMGs in their maps) and employ the \textsc{stacker} library
developed for use in \textsc{casa} to stack the visibilities
\citep{lindroos15}. The resulting stacked visibilities are then imaged using \textsc{casa}.  
Since the SMGs in our sample have a range of flux and signal-to-noise ratio, we stack the data
weighted by 1\,/\,$\sigma^2$.
From the stacked visibilities weighted by 1\,/\,$\sigma^2$, we measure
the flux as a function of radius, and show the resulting surface
brightness profile as a function of radius in Fig.~\ref{fig:stack}.  This figure
shows a resolved, high surface brightness central region, but we also
clearly detect faint and extended emission on $\geq 0\farcs5$ scales,
with an integrated flux that is $\sim10$\% of the total flux.

To assess the sensitivity of the derived properties of the
extended component on details of the data processing, we also
derive a stacked profile by combining the individual image-plane
maps of the SMGs for comparison. We extract a 10$''\times$\,10$''$ thumbnail
centred on each SMG, and then average the thumbnails,
weighted by 1\,/\,$\sigma^2$.  We then again extract the surface brightness
profile from the stacked map and overlay this on to the profile
created from the $uv$-stack in Fig.~\ref{fig:stack}.  We apply the same procedure
to the calibrator, and also overlay this in the same figure.  The
image-plane- and $uv$-derived stacks are well matched, with both
showing the same extended emission on $\geq 0\farcs5$ scales. 
 We stress that the profile of the (point-source) calibrator, which we scale
to the same peak surface brightness, is much narrower than the compact
emission and lacks the faint emission halo we see in the SMGs.
To test the that this extended emission is not due to weak side-lobes in the individual maps, we use {\sc casa} to  simulate a source with an $n=1$ light profile and a size and \ar\ equal to that of our sample, but 10 times brighter. This allows us to test if the extended emission we see in the stacked imaged is due to weak emission on large radii from a compact source. This test show that a compact source with \Reff$ = 0\farcs11$ observed at \res\ resolution does not show emission on $\geq0\farcs5$ scale. This imply that the extended emission we detect in the stacked image is indeed likely to be caused by a second component.

To characterise the surface brightness profile, we fit a two-component
model, including an inner and an outer \sersic\ profile each with
$n=1$. 
For the compact component we fix \Reff\ to the value derived above, $0\farcs10\pm0\farcs04$, 
and for the extended component we adopt \Reff\ $\sim0\farcs5$. 
The extended component accounts for  $13\pm1$\% of the total emission.
This suggest that the SMGs generally comprise a centrally concentrated
starburst which accounts for $\sim90$\% of the total dust continuum
flux density, with an extended star formation component on scales
similar to that seen in the rest-frame UV/optical (as seen by \textit{HST}).   
The transition between the compact and extended
components occur at $0\farcs15\pm0\farcs01$ ($\sim1.2$\,kpc at $z\sim 3$),
and the luminosity-weighted average effective radius of the two components, which provide the most appropriate "size" for the whole systems in $0\farcs15\pm0\farcs05$ corresponding to $1.2\pm0.1$\,kpc.
However, in terms of relative surface brightness -- the extended
component has a peak surface brightness (and hence implied dust
mass surface density) which is around two orders of magnitude lower
than the compact component.
\section{Discussion}\label{sec:dis}

% late type HUBBLE_TYPE_CODE==15 ang GALMAG_g=17-18
% early type HUBBLE_TYPE_CODE==12 GALMAG_b<18

%%%%%%%%%%%%%%%%%%%%%%%%%%%%%%%%%%%%%%%%%%%%%%%%%%
%
% Figure 8
%
\begin{figure*}
\centering
\includegraphics[trim= 0cm 0cm 0cm 0cm, clip=true,scale=0.85,angle=0]{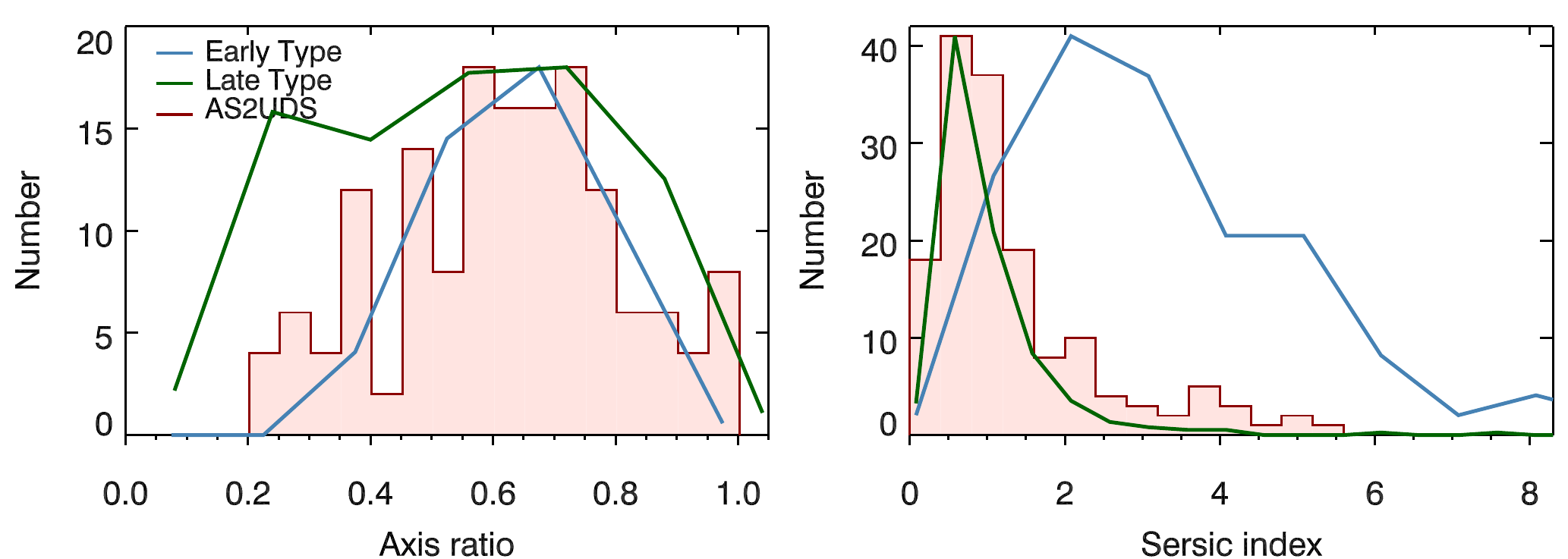}
\caption{{\small A comparison of the distribution of \ar s and \sersic\ indices for \num\ SMGs,
   to those measured at rest-frame optical wavelengths for
   morphologically-classified late-type discs and early-type spheroidal galaxies  from the GAMA survey \citep{kelvin14}. 
\textit{Left:} Axis ratio distribution of our SMGs compared to that of
the late-type disc and spheroids, with the distribution scaled to the peak of
our SMG distribution. A KS test shows that our SMG distribution is
most similar to that of the spheroidal galaxies, with a $30$\%
chance that these are selected from the same parent sample (compared
to $0$\% match to the late-type discs). This supports the validity of
the triaxial model for the morphologies of the SMGs proposed in \S~\ref{sec:tria}. 
\textit{Right:} \sersic\ index distribution of our SMGs compared to
that of late-type discs and spheroids, now showing a stronger similarity between the
SMGs and the late-type disc galaxies, rather than the spheroids (which
are rejected as a match based on KS test
returning a $0$\% probability that they are selected from the same
parent sample). These two plots thus suggest that the typical \cont\ dust
emission of our SMGs has a triaxial morphology, but an exponential
surface profile -- these characteristics are seen for central bars in disc galaxies.  
}}
\label{fig:gama}
\end{figure*}

\subsection{Discs, spheroids or bars?}

To provide a qualitative context for the structural properties of the SMGs in
our sample, we compare in Fig.~\ref{fig:gama}  the observed \ar\ and \sersic\ index
distributions  from the \cont\ observations of the SMGs to those
derived in the $g$-band  for the stellar light distributions of
a morphologically
classified sample of low-redshift galaxies 
from the GAMA survey \citep{kelvin14}. 
Consistent with the results from our modelling in the previous
section, Fig.~\ref{fig:gama} shows that 
the apparent \ar\ distribution of disc galaxies is not a good
match to that observed for the SMGs, with a 0\% chance that they are
drawn from the same parent population.
Whereas the SMG distribution is  better matched  to that of the
spheroids, with a KS test returning a 30\% chance of the two samples
sharing the same parent population.   

However, Fig.~\ref{fig:gama} also shows that
spheroids have light profiles which are better described by high \sersic\ indexes. 
This is not the case for the SMGs, which have a median \sersic\ index
of $n=1.00\pm0.12$, with a distribution which differs
significantly from the spheroids (a KS test returns a 0\% probability
that
the two distributions are from the same parent sample).   Whereas
the \sersic\ index distribution for  late-type disks is a much closer
match to that seen for the SMGs, with a strong peak at $n\sim 1$,
and a tail to higher $n$. 

This combination of apparently triaxial structures with exponential
surface brightness profiles resembles that seen in the 
central bars of barred spiral galaxies  \citep[e.g.][]{seigar98}.   This
interpretation
of the morphology of the dust emission in SMGs was first 
suggested by \cite{hodge19}.  They re-observed   
six of the ALESS SMGs from \cite{hodge16} with ALMA in
deep integrations at 0\farcs07
resolution and identified complex structures which were
unresolved in their earlier 0\farcs15 observations (comparable to the
data
analysed here).  In their higher resolution and deeper maps, \cite{hodge19} find
symmetric clump-like structures bracketing elongated nuclear emission. They
interpret these morphologies as representing bars in galaxies where the
``clump''-like structures are formed through orbit crowding or
star-forming rings. \cite{hodge19} find a ratio of the diameters of
the bar-to-ring structures  of $1.9\pm0.3$ similar to that seen for
these components in local barred galaxies. 
Our results on a larger statistical sample, while lacking the
resolution and depth to directly detect these features, have
structural properties consistent with the suggestion that much of the dust continuum emission from SMGs
arises in bar-like structures in their central regions.

%
% Figure 9
%
\begin{figure*}
\centering \includegraphics[trim= 0cm 0cm 0.cm 0cm,
clip=true,scale=0.84,angle=0]{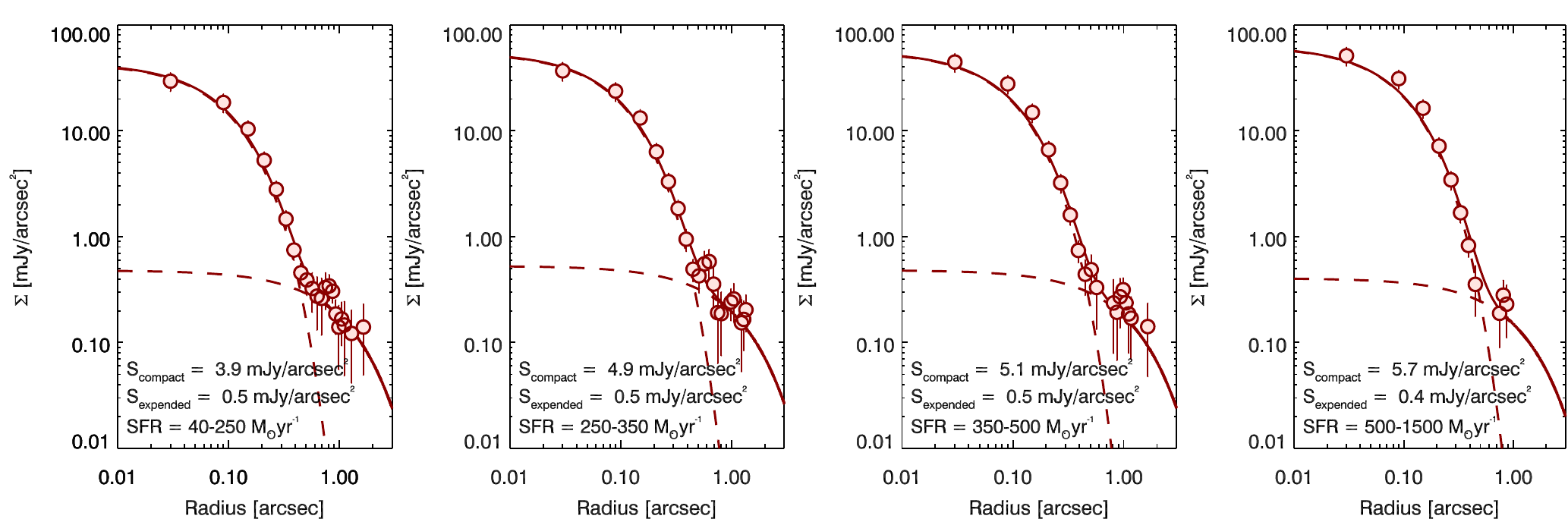}
\caption{{\small Surface brightness as a function of radius for the
  image stacks split into bins of star-formation rates of 40--250,
  250--350, 350--500 and 500--1500\,M$_{\odot}$\,yr$^{-1}$. Both
  stacks in the image plane and visibility plane return similar
  profiles, which are composed of a compact and an extended
  component (we therefore only show the image stack). We fit double
  exponential profiles and find that the peak flux of the extended
  component is close to constant at a mean of $0.47\pm0.03$\,mJy for
  the four stacks, while the compact component becomes brighter with
  increasing star-formation rate. }}
\label{fig:stacks_bins}
\end{figure*} 

\subsection{What is the extended dust component?}

Our analysis of the stacked profiles of the SMGs in our sample
indicates the presence of a spatially-extended component with a 
peak surface brightness which is roughly two orders of magnitude
fainter than the compact component detected in the individual sources,
and which contributes $\sim 10$\% to the total flux densities.

\begin{figure}
\centering
\includegraphics[trim= 0.4cm 0cm 0.1cm 0.1cm, clip=true,scale=0.84,angle=0]{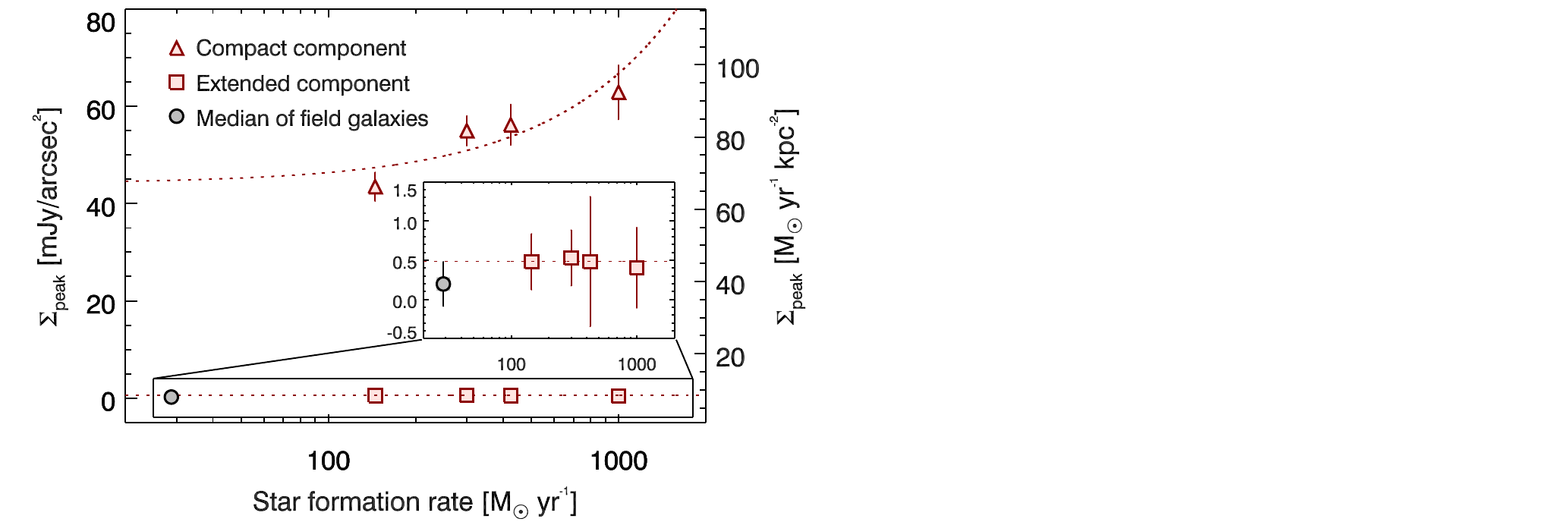}
\caption{{\small Fitted peak surface brightness as a function of the
  star-formation rate for the compact and extended components (in Fig.~\ref{fig:stacks_bins}). We see
  that the peak surface brightness for the extended component is roughly constant,
  while the brightness of the compact component increases as a
  function of star-formation rate. 
This is illustrated by the linear fits. We
  compare the extended components to the 
surface brightnesses of faint dusty galaxies in the
  {\it Hubble}
  Frontier Fields, and find a comparable median surface
  brightness. Hence the extended component we find in the SMGs are comparable in dust surface density to "normal" star-forming field galaxies.}}
\label{fig:excom}
\end{figure}

To investigate the relationship between the compact and
extended dust continuum emission, we split our sample into 
four bins of star-formation rate (with equal numbers of SMGs in
each bin) and stack the maps of these sources in
the image plane.  We show the resulting  surface brightness
profiles for the four independent subsamples  in Fig.~\ref{fig:stacks_bins}. Each of
these profiles shows both a compact and an extended component.
We fit these with the same  double \sersic\ model used above to derive the fraction of luminosity in the compact and
extended components.   We then plot this ratio as a function of star-formation rate in Fig.~\ref{fig:excom}.  This figure demonstrates that
the luminosity density of the extended emission remains approximately
constant, despite the luminosity density of the compact component
increasing by a factor of 50\% (over a range of a factor $\sim6$ in
total star-formation rate), suggesting that the star-formation surface
density in the compact and extended regions are decoupled.

We also compare our measurements of the
extended component with the median value of a sample of
faint field galaxies in strongly lensed cluster in the {\it Hubble} Frontier Fields
Survey \citep{gonzalez-lopez17, laporte17}.  These galaxies at
$z\simeq1.0$--$2.9$ are detected with ALMA at 1.1\,mm and represent
more typical star-forming field
galaxies with star-formation rates of $10$--$100$\,M$_{\odot}$\,yr$^{-1}$.
Interestingly, the median value for these faint field galaxies 
show similar surface brightness and extent to the extended components
seen in the SMGs, and suggests that the extended star formation we
detect in SMGs has similar star-formation surface density to that in
``normal'' star-forming galaxies in the field.  In contrast, the intense star-formation surface density in the
compact component is significantly higher.

We also wish to understand the relationship of the extended structure
we have uncovered at \cont\ to the other baryonic components of these
systems:  the (unobscured) stellar emission and the cool molecular gas.
To demonstrate how the observed near-infrared emission (rest-frame
UV/optical at $z\sim 3$) compares to the \cont\ emission, in
Fig.~\ref{fig:stack} we overlay the surface brightness profile
of the dust continuum emission and the near-infrared from \emph{HST}
(in this figure, we have scaled the surface brightness profiles to the
same integral).  We also overlay the gas emission as inferred from the
molecular $^{12}$CO(3--2) emission from four SMGs from ALESS
\citep{rivera18}.  This low-$J$ CO transition is expected to arise
from   material in the interstellar medium which has low to moderate critical
densities. This means its extent should trace the bulk of the underlying cool gas reservoir in
these systems.     As can be seen in Fig.~\ref{fig:stack}, the
extended emission seen at \cont\ from our stack seems to match
the spatial extent of the sources as seen
in the near-infrared and also in the low-$J$ CO from the molecular gas.
Since the extended \cont\ dust continuum component follows the same
profile as the molecular gas and (less obscured) stellar emission,
this suggests that the extended component traces a halo or outer disc in
these galaxies, which are dominated by stars and with much lower star-formation surface densities and obscuration than the central
starbursts.

Similar two-component profiles to that we see in our \cont\ stacking analysis have also been observed through stellar-mass surface-density profiles in high-resolution hydrodynamical simulations of merging high-redshift massive starburst discs with properties similar to those of SMGs \citep{hopkins13}.
The merging galaxies in these simulations are initially disc-dominated, but form nuclear bulges (on kpc scales) dominated by in situ star formation fuelled by gas driven to the centre by strong torques \citep{mihos94, hopkins08}. 
Using dynamical arguments \cite{hopkins09} suggest that because gas can dissipate energy, it can efficiently lose its angular momentum and rapidly fall into the centre in a merger event. 
This results in a concentrated starburst event seen in, for example, nearby merging ULIRGs and recent merger remnants \citep[e.g.][]{scoville86, sargent87, sargent89, kormendy92, rothberg04}. This process builds a clear bulge in the centre of the merger remnant. 

The extended component or ``envelope'' in this simulation is dominated by stars formed before the merger event and gas at large radii with significant angular momentum. This gas does not lose its angular momentum in the merger and re-forms a disc as the remnant relaxes. 
The survival (or re-formation) of the disc is therefore dependent on  how much gas loses its angular momentum \citep{hopkins09}. 
If all the gas in a merger event efficiently loses its angular
momentum, it would all be consumed in the nuclear starburst, and no
gas would be left to re-form a disc. However, high gas fractions have
been shown to be inefficient at losing their angular momentum, leading
to the fraction of gas available to fuel the central starburst scaling
sub-linearly with the gas fraction, and so leaving gas to re-form a disc \citep{hopkins13}.

In the framework of the model developed by \cite{hopkins13} we can
also investigate the physical nature of the SMGs and their triggering.
Besides considering different models with feedback and effective
equations of state, \cite{hopkins13} also consider the influence of
prograde versus retrograde mergers of disc galaxies. The different
feedback models and the relative angular momentum vectors  of the discs have little influence on the stellar mass profile of the remnant. 
However, \cite{hopkins13} also show that the relative angular momentum
vectors and orbit
of the merging components  has an influence on the time scales with which the merger remnant evolves. 
A prograde merger develops a morphology with a nuclear starburst and an envelope after $\sim1$\,Gyr, while the morphology of a retrograde merger after $\sim1$\,Gyr still shows two separate disc galaxies. 
This suggest that if major-merger events are the cause of the nuclear starburst event that we observe for our SMGs, a large fraction will have to be remnants of, or late stage, prograde mergers. 
If they were retrograde then the median redshift of our sample of $z\sim2.9$ this means that the
two merging disc galaxies would have to have appeared as highly star-forming
systems at redshift $\gtrsim5$. 
The redshift distribution of SMGs \citep{chapman05, weiss13,
 simpson14, ugne19} has been shown to peak at
$z\sim2.5$--$3.5$ and to have a tail out to redshifts $z\sim7$,
meaning that starburst galaxies at $z\gtrsim5$ do occur, but are rare
in \cont-selected samples.  Hence we suggest that SMGs at $z\sim2$--$3$ are more likely to be late-stage mergers, rather than merger remnants.

%
% Figure 10
%

%
%
%

%%%%%%%%%%%%%%%%%%%%%%%%%%%%%%%%%%%%%%%%%%%%%%%%%%
%
%
%
\section{Summary and Conclusions}\label{sec:con}
We analyse the dust continuum morphologies and light profiles of \num\
well-detected (S/N\,$>8$) SMGs observed with ALMA at \res\ ($\sim1$\,kpc) resolution. 
We fit both Gaussians (in the visibility and image plane) and free and $n=1$
\sersic\ models (in the image plane), and measure the effective
radii, \ar\ and \sersic\ indices for the individual sources.  We also
stack (again in both the visibility and image planes) the \cont\
emission for the whole sample and selected subsets to trace fainter
and more extended emission around these systems.  
%Our analysis of the dust morphologies of SMGs has shown that their \ar\ distribution is best described by non-asymmetric (triaxial) morphologies and the \sersic\ index distribution suggest that their brightness profiles are best described by a \sersic\ index of $n\sim1$ in the central  $\sim 1$--2\,kpc. 
%These are characteristics of bars in galaxies, which drive gas from the outer disc to the centre, thereby fuelling the central star-formation. 
%We also find that while this bar-like component is detected on a case-by-case basis, a much fainter, extended outer halo is only detected by stacking. 
%The stacked image reveals a two-component profile with a compact and extended component, tracing respectively a bar and the disc or envelope. 
Our main conclusions are:

$\bullet$ The median effective radius for SMGs in our sample is $0\farcs10\pm0\farcs04$. 
Accounting for the extended dust component we find, we derive a flux-weighted effective radius of $0\farcs15\pm0\farcs05$ or $1.2\pm0.4$\,kpc at $z\sim3$.    
This in consistent with estimates of the sub-millimetre sizes of SMGs
from earlier smaller-scale studies. %, which derived effective radii of $0\farcs15\pm0\farcs03$--$0\farcs13\pm0\farcs02$\citep{simpson15b, hodge16, ikarashi17, gullberg18}.
We show that there is a rough correlation between the \cont\ and
observed
$K$-band sizes of SMGs in our sample, with the \cont\ sizes
being on average $2.2\pm 0.2$ times smaller.

$\bullet$ The effective radii of SMGs in our sample show a very weak
decline with increasing redshifts and a similarly marginal increase with \flux . Using the physical properties of
our sample from \cite{ugne19}  and
assuming a simple source geometry we expect that the typical apparent source
size would decrease slightly at higher redshifts, although this
evolution would be countered by optical depth effects.   
The weak decline in size we see with redshift suggests that
the physical sizes of SMGs do not evolve strongly with redshift 
and the lack of variation in size with \flux\ indicates that the more
luminous systems are likely to exhibit higher pressures
in their interstellar medium. 

$\bullet$ We find that the apparent \ar\ distribution  of the SMGs is best described by non-axisymmetric morphologies (triaxial) and the \sersic\ index distribution has a median of $n=1.0\pm0.1$. 
By comparing these distributions with those of disc and spheroid
galaxies, the \ar\ distribution of SMGs is most similar to those of
spheroid galaxies, while their \sersic\ index distribution is most
similar to that of disc galaxies. This combination of exponential
surface brightness profiles and triaxial structures are the
characteristics of bars in galaxies. Higher resolution and deeper
observations of six SMGs by \cite{hodge19} have identified potential
bar and ring structures in those galaxies and we  therefore suggest
that the statistical properties of the SMGs in our sample point to
bars being a ubiquitous feature of bright SMGs.  

$\bullet$ We stack our SMGs in both the image and visibility planes
and find that the continuum emission profiles are composed of not only
the  compact component we have directly detected, but also a much
lower surface brightness,  extended component.  
The extended component accounts for $13\pm1$\% of the total emission
and has a scale size of $\sim0\farcs5$ ($\sim4$\,kpc) 
Comparing with stacked CO(3--2) and \textit{HST} imaging of samples of
SMGs, we see that the extended component is
comparable in size to the low-$J$ CO molecular gas and stellar
distributions. 
We conclude that it is likely that the extended component seen in the stacked \cont\ maps traces a surrounding disc or envelope around the central, compact far-infrared luminous starburst.

$\bullet$ By stacking the \cont\ maps in bins of star-formation rate
we find that the size and luminosity of the extended component is
roughly constant with total star-formation rate, while the compact component becomes brighter. This suggests that the star formation taking place in the compact component is broadly decoupled from the star formation taking place in the extended component.
\\\\
We have studied a large sample of SMGs using moderate resolution ALMA data and find that the morphologies observed at \res\ resolution are best described by bars in galaxies. 
However, to confirm this requires deeper observations to detect the extended components in individual maps, higher resolution imaging ($\sim0\farcs08$) to show that the \cont\ dust continuum trace bars structures, and ideally dynamical gas measurements of the extended component to determine if this has an order rotational motion as seen for disc or more chaotic as would be expected from a merger event. 

\section*{Acknowledgements}
We thank the anonymous referee for their helpful thorough reading of the manuscript, and suggestions that improved the paper.
BG and IRS acknowledge support from the ERC Advanced Investigator program DUSTYGAL (321334).
BG, IRS and AMS acknowledge financial support from an STFC grant (ST/P000541/1).  
JLW acknowledges support from an STFC Ernest Rutherford Fellowship (ST/P004784/1 and ST/P004784/2).
MJM acknowledges the support of the National Science Centre, Poland
through the SONATA BIS grant 2018/30/E/ST9/00208
The ALMA data used in this paper were obtained under programs ADS/JAO.ALMA\#2015.1.01528.S. ALMA is a partnership of ESO (representing its member states), NSF (USA) and NINS (Japan), together with NRC (Canada) and NSC and ASIAA (Taiwan), in cooperation with the Republic of Chile. The Joint ALMA Observatory is operated by ESO, AUI/NRAO, and NAOJ.

%%%%%%%%%%%%%%%%%%%%%%%%%%%%%%%%%%%%%%%%%%%%%%%%%%



\bibliographystyle{mnras}
\bibliography{AS2UDS}
\section*{Appendix A}

\begin{figure*}
\centering
\includegraphics[trim= 0.1cm 0.1cm 0.1cm 0.1cm, clip=true,scale=0.75,angle=0]{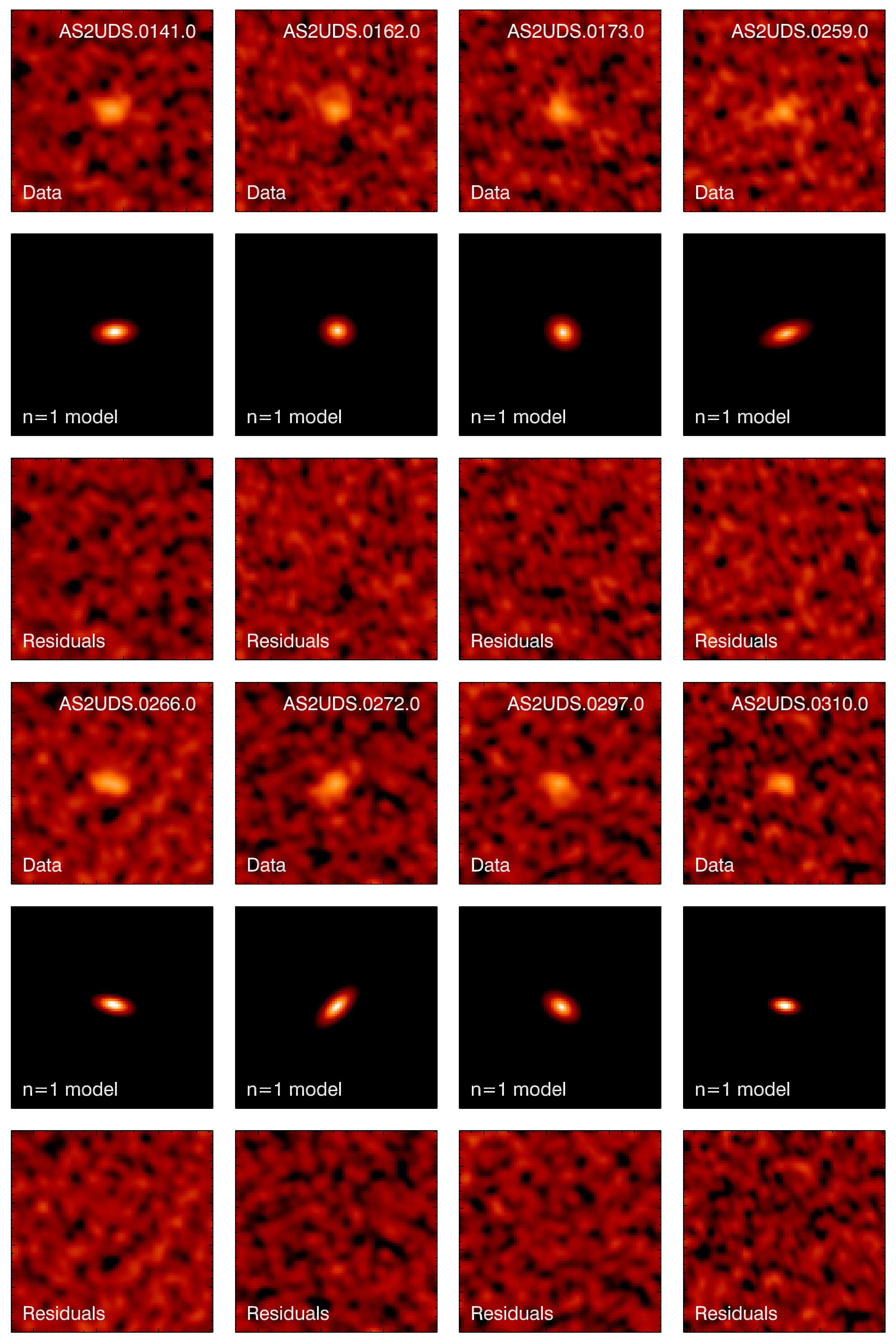}
\caption{{\small Examples of \sersic\ fits with $n=1$ to the eight examples shown in Fig.~\ref{fig:maps_hst}. The rows show the data, model and the residuals.}}
\label{fig:maps_models2}
\end{figure*} 

\onecolumn
\begin{longtable}{l c c c c c c c}
\hline\hline
Name & photometric redshift & SFR                     & $R_e({n=1})$ & $R_e(\text{free})$ & Axis ratio & $\theta$ & $n$ \\
     &                      &  M$_{\odot}$\,yr$^{-1}$ &[arcsec]            &   [arcsec]  &            & [degree] &     \\
\hline
AS2UDS\ 0009.0  &  $2.635 ^{2.319 }_{2.951}$  & 646  &$0.15\pm0.01$&  $ 0.18\pm 0.01$&  $0.60\pm0.03$&  $ 66\pm  1$&  $0.7\pm0.0$ \\
AS2UDS\ 0010.0  &  $3.945 ^{3.625 }_{4.135}$  &1122  &$0.15\pm0.02$&  $ 0.48\pm 0.01$&  $0.99\pm0.02$&  $ 36\pm259$&  $5.2\pm0.9$ \\
AS2UDS\ 0011.0  &  $3.735 ^{2.835 }_{4.805}$  & 741  &$0.10\pm0.01$&  $ 0.12\pm 0.00$&  $0.73\pm0.02$&  $130\pm  2$&  $0.8\pm0.1$ \\
AS2UDS\ 0012.0  &  $2.435 ^{2.235 }_{2.635}$  & 355  &$0.16\pm0.02$&  $ 0.19\pm 0.01$&  $0.60\pm0.04$&  $ 52\pm  2$&  $0.7\pm0.1$ \\
AS2UDS\ 0014.0  &  $3.685 ^{3.565 }_{3.905}$  & 759  &$0.10\pm0.01$&  $ 0.12\pm 0.00$&  $0.62\pm0.02$&  $ 50\pm  1$&  $0.8\pm0.0$ \\
AS2UDS\ 0016.0  &  $2.755 ^{2.605 }_{2.795}$  & 724  &$0.07\pm0.01$&  $ 0.10\pm 0.00$&  $0.58\pm0.01$&  $ 59\pm  1$&  $0.6\pm0.1$ \\
AS2UDS\ 0018.0  &  $3.615 ^{3.505 }_{3.735}$  & 575  &$0.08\pm0.01$&  $ 0.13\pm 0.01$&  $0.33\pm0.03$&  $159\pm  1$&  $0.7\pm0.1$ \\
AS2UDS\ 0026.0  &  $3.825 ^{3.395 }_{4.845}$  & 490  &$0.15\pm0.01$&  $ 0.23\pm 0.01$&  $0.53\pm0.03$&  $ 61\pm  1$&  $3.8\pm3.0$ \\
AS2UDS\ 0028.0  &  $3.105 ^{2.495 }_{3.615}$  & 589  &$0.07\pm0.01$&  $ 0.12\pm 0.00$&  $0.38\pm0.02$&  $ 42\pm  1$&  $0.9\pm0.1$ \\
AS2UDS\ 0028.0  &  $3.105 ^{2.495 }_{3.615}$  & 589  &$0.07\pm0.02$&  $ 0.12\pm 0.00$&  $0.38\pm0.05$&  $ 42\pm  1$&  $0.9\pm0.1$ \\
AS2UDS\ 0029.0  &  $1.945 ^{1.615 }_{2.255}$  & 550  &$0.09\pm0.02$&  $ 0.13\pm 0.01$&  $0.82\pm0.03$&  $149\pm  4$&  $2.7\pm0.3$ \\
AS2UDS\ 0030.0  &  $3.195 ^{2.975 }_{3.815}$  & 407  &$0.06\pm0.01$&  $ 0.14\pm 0.01$&  $0.30\pm0.03$&  $ 72\pm  1$&  $2.5\pm0.3$ \\
AS2UDS\ 0031.0  &  $4.875 ^{3.935 }_{5.185}$  &2188  &$0.11\pm0.01$&  $ 0.13\pm 0.00$&  $0.75\pm0.02$&  $ 70\pm  3$&  $0.9\pm0.1$ \\
AS2UDS\ 0032.0  &  $2.725 ^{2.605 }_{2.965}$  & 229  &$0.10\pm0.01$&  $ 0.12\pm 0.00$&  $0.73\pm0.02$&  $101\pm  2$&  $1.1\pm0.1$ \\
AS2UDS\ 0035.0  &  $1.615 ^{1.421 }_{1.809}$  & 355  &$0.09\pm0.03$&  $ 0.10\pm 0.01$&  $0.87\pm0.04$&  $ 16\pm 10$&  $0.9\pm0.1$ \\
AS2UDS\ 0038.0  &  $2.575 ^{2.485 }_{2.625}$  & 525  &$0.08\pm0.01$&  $ 0.11\pm 0.00$&  $0.57\pm0.02$&  $ 42\pm  2$&  $1.4\pm0.1$ \\
AS2UDS\ 0042.0  &  $3.505 ^{3.035 }_{4.445}$  & 389  &$0.08\pm0.01$&  $ 0.12\pm 0.01$&  $0.40\pm0.03$&  $157\pm  1$&  $1.3\pm0.1$ \\
AS2UDS\ 0046.0  &  $3.655 ^{3.385 }_{4.005}$  & 501  &$0.09\pm0.01$&  $ 0.11\pm 0.00$&  $0.52\pm0.02$&  $117\pm  1$&  $0.3\pm0.0$ \\
AS2UDS\ 0046.0  &  $3.655 ^{3.385 }_{4.005}$  & 501  &$0.06\pm0.02$&  $ 0.11\pm 0.00$&  $0.51\pm0.04$&  $ 72\pm  4$&  $0.3\pm0.0$ \\
AS2UDS\ 0048.0  &  $3.065 ^{2.925 }_{3.365}$  &1122  &$0.09\pm0.01$&  $ 0.11\pm 0.00$&  $0.67\pm0.02$&  $145\pm  3$&  $0.6\pm0.1$ \\
AS2UDS\ 0049.0  &  $2.645 ^{2.515 }_{2.815}$  & 347  &$0.11\pm0.02$&  $ 0.16\pm 0.01$&  $0.48\pm0.04$&  $ 62\pm  2$&  $1.0\pm0.1$ \\
AS2UDS\ 0050.0  &  $3.695 ^{3.205 }_{4.515}$  & 562  &$0.11\pm0.01$&  $ 0.11\pm 0.00$&  $0.97\pm0.02$&  $ 74\pm 30$&  $1.1\pm0.1$ \\
AS2UDS\ 0054.0  &  $2.715 ^{2.595 }_{2.785}$  & 324  &$0.08\pm0.01$&  $ 0.13\pm 0.00$&  $0.49\pm0.02$&  $ 29\pm  1$&  $1.7\pm0.2$ \\
AS2UDS\ 0055.0  &  $2.545 ^{2.485 }_{2.605}$  & 123  &$0.12\pm0.02$&  $ 0.17\pm 0.01$&  $0.78\pm0.04$&  $174\pm  6$&  $1.9\pm0.2$ \\
AS2UDS\ 0058.0  &  $3.855 ^{3.255 }_{4.535}$  & 692  &$0.10\pm0.02$&  $ 0.10\pm 0.00$&  $0.99\pm0.02$&  $108\pm 93$&  $1.2\pm0.1$ \\
AS2UDS\ 0059.0  &  $3.205 ^{2.975 }_{3.755}$  & 575  &$0.11\pm0.02$&  $ 0.14\pm 0.01$&  $0.52\pm0.03$&  $ 90\pm  1$&  $0.3\pm0.0$ \\
AS2UDS\ 0061.0  &  $3.635 ^{3.245 }_{4.395}$  & 398  &$0.08\pm0.02$&  $ 0.13\pm 0.01$&  $0.35\pm0.05$&  $115\pm  1$&  $0.7\pm0.1$ \\
AS2UDS\ 0062.0  &  $4.835 ^{4.745 }_{4.955}$  &1413  &$0.09\pm0.01$&  $ 0.10\pm 0.00$&  $0.78\pm0.02$&  $ 90\pm  5$&  $0.9\pm0.1$ \\
AS2UDS\ 0063.0  &  $4.985 ^{3.125 }_{5.375}$  &1445  &$0.11\pm0.01$&  $ 0.11\pm 0.00$&  $0.93\pm0.02$&  $ 69\pm 13$&  $1.1\pm0.1$ \\
AS2UDS\ 0064.0  &  $4.145 ^{3.835 }_{4.635}$  & 589  &$0.09\pm0.01$&  $ 0.12\pm 0.00$&  $0.52\pm0.03$&  $ 44\pm  2$&  $0.5\pm0.1$ \\
AS2UDS\ 0066.0  &  $2.035 ^{1.755 }_{2.115}$  & 389  &$0.09\pm0.01$&  $ 0.12\pm 0.00$&  $0.58\pm0.02$&  $ 85\pm  2$&  $0.7\pm0.1$ \\
AS2UDS\ 0067.0  &  $1.455 ^{1.435 }_{1.475}$  & 447  &$0.09\pm0.01$&  $ 0.12\pm 0.00$&  $0.68\pm0.02$&  $  1\pm  2$&  $1.8\pm0.2$ \\
AS2UDS\ 0072.0  &  $2.875 ^{2.755 }_{2.995}$  & 741  &$0.07\pm0.01$&  $ 0.09\pm 0.00$&  $0.66\pm0.02$&  $ 57\pm  3$&  $1.4\pm0.2$ \\
AS2UDS\ 0073.0  &  $2.455 ^{2.425 }_{2.465}$  & 891  &$0.12\pm0.01$&  $ 0.13\pm 0.00$&  $0.98\pm0.02$&  $ 90\pm 70$&  $1.4\pm0.1$ \\
AS2UDS\ 0074.0  &  $2.965 ^{2.609 }_{3.321}$  & 245  &$0.12\pm0.02$&  $ 0.12\pm 0.01$&  $0.65\pm0.04$&  $139\pm  3$&  $0.2\pm0.1$ \\
AS2UDS\ 0075.0  &  $2.545 ^{2.345 }_{2.625}$  & 676  &$0.15\pm0.02$&  $ 0.14\pm 0.01$&  $0.93\pm0.03$&  $  6\pm 12$&  $0.6\pm0.0$ \\
AS2UDS\ 0076.0  &  $3.615 ^{3.505 }_{3.675}$  & 537  &$0.14\pm0.02$&  $ 0.14\pm 0.01$&  $0.94\pm0.03$&  $ 36\pm 20$&  $0.7\pm0.1$ \\
AS2UDS\ 0079.0  &  $3.755 ^{3.185 }_{4.945}$  & 468  &$0.10\pm0.01$&  $ 0.14\pm 0.00$&  $0.68\pm0.02$&  $ 81\pm  2$&  $1.5\pm0.1$ \\
AS2UDS\ 0083.0  &  $3.485 ^{3.165 }_{3.965}$  & 724  &$0.07\pm0.01$&  $ 0.08\pm 0.00$&  $0.71\pm0.02$&  $164\pm  5$&  $0.4\pm0.1$ \\
AS2UDS\ 0087.0  &  $4.005 ^{3.405 }_{4.565}$  & 347  &$0.08\pm0.01$&  $ 0.10\pm 0.00$&  $0.63\pm0.02$&  $142\pm  3$&  $0.8\pm0.1$ \\
AS2UDS\ 0089.0  &  $3.785 ^{3.205 }_{4.515}$  & 631  &$0.06\pm0.01$&  $ 0.10\pm 0.00$&  $0.43\pm0.02$&  $141\pm  1$&  $1.2\pm0.1$ \\
AS2UDS\ 0090.0  &  $3.125 ^{2.385 }_{4.215}$  & 224  &$0.10\pm0.03$&  $ 0.11\pm 0.01$&  $0.59\pm0.05$&  $ 82\pm  3$&  $0.2\pm0.1$ \\
AS2UDS\ 0092.0  &  $3.875 ^{3.175 }_{5.495}$  & 372  &$0.08\pm0.01$&  $ 0.11\pm 0.00$&  $0.45\pm0.03$&  $ 63\pm  1$&  $0.5\pm0.1$ \\
AS2UDS\ 0093.0  &  $2.255 ^{1.645 }_{3.115}$  & 204  &$0.08\pm0.01$&  $ 0.13\pm 0.00$&  $0.51\pm0.02$&  $171\pm  1$&  $2.0\pm0.2$ \\
AS2UDS\ 0096.0  &  $3.005 ^{2.955 }_{3.045}$  & 724  &$0.09\pm0.02$&  $ 0.12\pm 0.01$&  $0.76\pm0.04$&  $ 24\pm  4$&  $1.9\pm0.2$ \\
AS2UDS\ 0098.0  &  $2.585 ^{2.545 }_{2.675}$  & 302  &$0.11\pm0.02$&  $ 0.13\pm 0.01$&  $0.65\pm0.03$&  $ 56\pm  3$&  $0.6\pm0.1$ \\
AS2UDS\ 0099.0  &  $2.935 ^{2.905 }_{2.965}$  & 724  &$0.12\pm0.02$&  $ 0.14\pm 0.01$&  $0.63\pm0.04$&  $ 95\pm  3$&  $0.6\pm0.1$ \\
AS2UDS\ 0100.0  &  $3.085 ^{2.875 }_{3.235}$  & 398  &$0.12\pm0.02$&  $ 0.15\pm 0.01$&  $0.63\pm0.04$&  $ 14\pm  3$&  $0.9\pm0.1$ \\
AS2UDS\ 0102.0  &  $2.255 ^{2.005 }_{2.475}$  &  91  &$0.06\pm0.02$&  $ 0.10\pm 0.01$&  $0.31\pm0.05$&  $130\pm  1$&  $0.4\pm0.1$ \\
AS2UDS\ 0103.0  &  $3.075 ^{3.055 }_{3.115}$  & 501  &$0.07\pm0.01$&  $ 0.10\pm 0.00$&  $0.39\pm0.03$&  $ 73\pm  1$&  $0.2\pm0.1$ \\
AS2UDS\ 0104.0  &  $5.845 ^{5.005 }_{6.525}$  & 646  &$0.15\pm0.02$&  $ 0.50\pm 0.02$&  $0.56\pm0.04$&  $115\pm  2$&  $3.5\pm0.5$ \\
AS2UDS\ 0107.0  &  $3.015 ^{2.965 }_{3.085}$  & 562  &$0.08\pm0.01$&  $ 0.10\pm 0.00$&  $0.76\pm0.02$&  $ 64\pm  3$&  $1.0\pm0.1$ \\
AS2UDS\ 0110.0  &  $3.505 ^{3.385 }_{3.585}$  &1202  &$0.10\pm0.02$&  $ 0.23\pm 0.01$&  $0.82\pm0.03$&  $ 25\pm  8$&  $4.2\pm0.8$ \\
AS2UDS\ 0112.0  &  $1.805 ^{1.335 }_{2.605}$  & 229  &$0.09\pm0.01$&  $ 0.11\pm 0.00$&  $0.71\pm0.02$&  $ 87\pm  3$&  $1.0\pm0.1$ \\
AS2UDS\ 0120.0  &  $3.445 ^{3.135 }_{3.915}$  & 479  &$0.08\pm0.02$&  $ 0.21\pm 0.02$&  $0.30\pm0.06$&  $ 90\pm  1$&  $2.2\pm0.2$ \\
AS2UDS\ 0124.0  &  $3.665 ^{3.045 }_{4.675}$  & 331  &$0.10\pm0.02$&  $ 0.12\pm 0.01$&  $0.68\pm0.03$&  $176\pm  3$&  $0.9\pm0.1$ \\
AS2UDS\ 0126.0  &  $2.575 ^{1.955 }_{3.315}$  & 676  &$0.09\pm0.02$&  $ 0.11\pm 0.01$&  $0.60\pm0.04$&  $141\pm  2$&  $0.7\pm0.1$ \\
AS2UDS\ 0128.0  &  $3.265 ^{2.955 }_{4.155}$  & 257  &$0.10\pm0.02$&  $ 0.15\pm 0.01$&  $0.50\pm0.04$&  $ 54\pm  2$&  $1.2\pm0.1$ \\
AS2UDS\ 0132.0  &  $3.075 ^{3.015 }_{3.215}$  & 372  &$0.07\pm0.01$&  $ 0.07\pm 0.00$&  $0.85\pm0.02$&  $ 55\pm  8$&  $1.4\pm0.2$ \\
AS2UDS\ 0135.0  &  $3.615 ^{3.175 }_{5.145}$  & 708  &$0.10\pm0.01$&  $ 0.17\pm 0.00$&  $0.77\pm0.02$&  $ 71\pm  3$&  $2.7\pm0.2$ \\
AS2UDS\ 0137.0  &  $2.615 ^{2.605 }_{2.625}$  & 933  &$0.05\pm0.01$&  $ 0.07\pm 0.00$&  $0.56\pm0.02$&  $116\pm  3$&  $1.0\pm0.2$ \\
AS2UDS\ 0141.0  &  $1.725 ^{1.695 }_{1.915}$  & 309  &$0.08\pm0.02$&  $ 0.11\pm 0.01$&  $0.55\pm0.03$&  $ 95\pm  2$&  $0.6\pm0.1$ \\
AS2UDS\ 0143.0  &  $2.545 ^{2.525 }_{2.565}$  & 631  &$0.07\pm0.02$&  $ 0.14\pm 0.01$&  $0.81\pm0.03$&  $ 58\pm  9$&  $4.7\pm1.3$ \\
AS2UDS\ 0146.0  &  $2.985 ^{2.795 }_{3.225}$  & 437  &$0.12\pm0.01$&  $ 0.18\pm 0.00$&  $0.80\pm0.02$&  $131\pm  4$&  $2.2\pm0.2$ \\
AS2UDS\ 0147.0  &  $2.175 ^{1.535 }_{2.825}$  & 200  &$0.06\pm0.01$&  $ 0.08\pm 0.00$&  $0.68\pm0.02$&  $ 54\pm  4$&  $0.8\pm0.2$ \\
AS2UDS\ 0149.0  &  $3.065 ^{2.975 }_{3.205}$  & 316  &$0.04\pm0.01$&  $ 0.09\pm 0.00$&  $0.21\pm0.04$&  $ 16\pm  1$&  $1.0\pm0.1$ \\
AS2UDS\ 0151.0  &  $2.995 ^{2.785 }_{3.175}$  & 501  &$0.08\pm0.01$&  $ 0.10\pm 0.00$&  $0.64\pm0.02$&  $164\pm  3$&  $0.7\pm0.1$ \\
AS2UDS\ 0158.0  &  $2.965 ^{2.585 }_{3.115}$  & 398  &$0.08\pm0.01$&  $ 0.08\pm 0.00$&  $1.00\pm0.02$&  $144\pm***$&  $0.5\pm0.1$ \\
AS2UDS\ 0162.0  &  $3.765 ^{3.155 }_{5.165}$  & 355  &$0.11\pm0.02$&  $ 0.12\pm 0.01$&  $0.87\pm0.03$&  $ 79\pm 11$&  $0.7\pm0.1$ \\
AS2UDS\ 0164.0  &  $3.135 ^{2.665 }_{5.145}$  & 339  &$0.10\pm0.02$&  $ 0.14\pm 0.01$&  $0.78\pm0.04$&  $130\pm  6$&  $2.0\pm0.3$ \\
AS2UDS\ 0165.0  &  $2.255 ^{1.945 }_{2.565}$  & 603  &$0.04\pm0.01$&  $ 0.09\pm 0.00$&  $0.23\pm0.03$&  $114\pm  1$&  $0.7\pm0.1$ \\
AS2UDS\ 0166.0  &  $2.305 ^{2.295 }_{2.315}$  & 309  &$0.05\pm0.01$&  $ 0.09\pm 0.00$&  $0.33\pm0.02$&  $ 64\pm  1$&  $0.6\pm0.1$ \\
AS2UDS\ 0167.0  &  $2.685 ^{2.363 }_{3.007}$  & 479  &$0.09\pm0.02$&  $ 0.11\pm 0.01$&  $0.64\pm0.03$&  $154\pm  3$&  $0.8\pm0.1$ \\
AS2UDS\ 0169.0  &  $3.595 ^{3.255 }_{4.065}$  & 575  &$0.11\pm0.02$&  $ 0.21\pm 0.01$&  $0.52\pm0.05$&  $165\pm  2$&  $4.8\pm3.0$ \\
AS2UDS\ 0172.0  &  $3.535 ^{3.305 }_{3.735}$  &1096  &$0.10\pm0.02$&  $ 0.11\pm 0.00$&  $0.92\pm0.02$&  $159\pm 15$&  $0.9\pm0.1$ \\
AS2UDS\ 0173.0  &  $2.835 ^{2.645 }_{3.025}$  & 575  &$0.12\pm0.02$&  $ 0.12\pm 0.01$&  $0.87\pm0.03$&  $ 45\pm  9$&  $0.7\pm0.1$ \\
AS2UDS\ 0175.0  &  $2.085 ^{1.915 }_{2.145}$  & 479  &$0.14\pm0.03$&  $ 0.15\pm 0.01$&  $0.71\pm0.04$&  $  4\pm  4$&  $0.5\pm0.1$ \\
AS2UDS\ 0178.0  &  $3.195 ^{2.345 }_{4.215}$  & 389  &$0.08\pm0.02$&  $ 0.09\pm 0.01$&  $0.80\pm0.03$&  $119\pm  7$&  $0.6\pm0.1$ \\
AS2UDS\ 0182.0  &  $1.705 ^{1.685 }_{1.725}$  & 309  &$0.06\pm0.02$&  $ 0.08\pm 0.01$&  $0.72\pm0.03$&  $ 79\pm  7$&  $2.1\pm0.5$ \\
AS2UDS\ 0183.0  &  $3.245 ^{3.055 }_{3.355}$  & 398  &$0.06\pm0.01$&  $ 0.07\pm 0.00$&  $0.83\pm0.02$&  $ 97\pm  9$&  $1.1\pm0.2$ \\
AS2UDS\ 0185.0  &  $5.325 ^{3.235 }_{6.375}$  & 575  &$0.10\pm0.02$&  $ 0.11\pm 0.00$&  $0.87\pm0.02$&  $137\pm  8$&  $0.5\pm0.1$ \\
AS2UDS\ 0187.0  &  $1.945 ^{1.775 }_{2.135}$  & 200  &$0.08\pm0.02$&  $ 0.10\pm 0.01$&  $0.65\pm0.03$&  $116\pm  3$&  $0.3\pm0.1$ \\
AS2UDS\ 0192.0  &  $3.375 ^{3.185 }_{4.185}$  & 316  &$0.14\pm0.02$&  $ 0.15\pm 0.01$&  $0.96\pm0.03$&  $ 18\pm 27$&  $1.3\pm0.1$ \\
AS2UDS\ 0199.0  &  $2.985 ^{2.965 }_{3.005}$  & 407  &$0.05\pm0.02$&  $ 0.09\pm 0.01$&  $0.28\pm0.04$&  $168\pm  1$&  $0.9\pm0.1$ \\
AS2UDS\ 0203.0  &  $1.855 ^{1.632 }_{2.078}$  & 214  &$0.11\pm0.02$&  $ 0.15\pm 0.01$&  $0.58\pm0.04$&  $ 58\pm  2$&  $1.1\pm0.1$ \\
AS2UDS\ 0205.0  &  $3.535 ^{3.085 }_{4.335}$  & 389  &$0.08\pm0.01$&  $ 0.10\pm 0.00$&  $0.63\pm0.02$&  $ 25\pm  2$&  $1.2\pm0.1$ \\
AS2UDS\ 0209.0  &  $3.595 ^{3.585 }_{3.685}$  & 355  &$0.10\pm0.02$&  $ 0.20\pm 0.01$&  $0.67\pm0.03$&  $ 56\pm  4$&  $2.9\pm0.4$ \\
AS2UDS\ 0210.0  &  $2.275 ^{2.175 }_{2.395}$  & 214  &$0.09\pm0.01$&  $ 0.11\pm 0.00$&  $0.63\pm0.02$&  $161\pm  3$&  $1.2\pm0.1$ \\
AS2UDS\ 0212.0  &  $2.595 ^{2.575 }_{2.955}$  & 209  &$0.09\pm0.02$&  $ 0.11\pm 0.01$&  $0.58\pm0.04$&  $135\pm  3$&  $0.6\pm0.1$ \\
AS2UDS\ 0213.0  &  $2.825 ^{2.625 }_{3.025}$  & 347  &$0.12\pm0.02$&  $ 0.17\pm 0.01$&  $0.52\pm0.04$&  $ 40\pm  2$&  $1.4\pm0.1$ \\
AS2UDS\ 0218.0  &  $2.985 ^{2.655 }_{3.515}$  & 224  &$0.05\pm0.02$&  $ 0.22\pm 0.01$&  $0.24\pm0.06$&  $117\pm  1$&  $4.2\pm3.0$ \\
AS2UDS\ 0222.0  &  $3.665 ^{3.255 }_{4.275}$  & 776  &$0.05\pm0.01$&  $ 0.07\pm 0.00$&  $0.62\pm0.01$&  $ 22\pm  2$&  $3.6\pm0.7$ \\
AS2UDS\ 0225.0  &  $3.375 ^{2.895 }_{4.535}$  & 234  &$0.07\pm0.02$&  $ 0.10\pm 0.01$&  $0.45\pm0.03$&  $ 61\pm  2$&  $0.6\pm0.1$ \\
AS2UDS\ 0226.0  &  $2.225 ^{2.195 }_{2.255}$  & 209  &$0.06\pm0.02$&  $ 0.10\pm 0.01$&  $0.52\pm0.03$&  $ 12\pm  3$&  $2.3\pm0.5$ \\
AS2UDS\ 0231.0  &  $2.955 ^{2.875 }_{3.075}$  & 347  &$0.09\pm0.02$&  $ 0.19\pm 0.01$&  $0.33\pm0.07$&  $140\pm  1$&  $1.7\pm0.2$ \\
AS2UDS\ 0232.0  &  $2.525 ^{2.235 }_{2.705}$  & 240  &$0.08\pm0.01$&  $ 0.08\pm 0.00$&  $0.95\pm0.02$&  $110\pm 28$&  $1.1\pm0.2$ \\
AS2UDS\ 0235.0  &  $4.345 ^{3.395 }_{5.815}$  & 380  &$0.10\pm0.02$&  $ 0.12\pm 0.01$&  $0.67\pm0.04$&  $ 83\pm  4$&  $1.2\pm0.2$ \\
AS2UDS\ 0236.0  &  $3.945 ^{3.205 }_{4.875}$  & 288  &$0.07\pm0.02$&  $ 0.10\pm 0.01$&  $0.44\pm0.04$&  $138\pm  2$&  $0.4\pm0.1$ \\
AS2UDS\ 0238.0  &  $2.175 ^{2.105 }_{2.285}$  & 178  &$0.11\pm0.02$&  $ 0.10\pm 0.01$&  $0.99\pm0.03$&  $109\pm111$&  $0.7\pm0.1$ \\
AS2UDS\ 0243.0  &  $1.775 ^{1.765 }_{1.785}$  &1072  &$0.06\pm0.02$&  $ 0.07\pm 0.01$&  $0.66\pm0.03$&  $104\pm  5$&  $0.2\pm0.2$ \\
AS2UDS\ 0255.0  &  $2.215 ^{1.949 }_{2.481}$  & 132  &$0.10\pm0.02$&  $ 0.19\pm 0.01$&  $0.57\pm0.04$&  $101\pm  3$&  $3.9\pm3.0$ \\
AS2UDS\ 0259.0  &  $1.855 ^{1.815 }_{1.895}$  & 363  &$0.09\pm0.02$&  $ 0.19\pm 0.01$&  $0.47\pm0.04$&  $109\pm  2$&  $3.5\pm3.0$ \\
AS2UDS\ 0265.0  &  $2.295 ^{2.185 }_{2.325}$  & 245  &$0.04\pm0.01$&  $ 0.06\pm 0.00$&  $0.37\pm0.02$&  $103\pm  2$&  $1.1\pm0.2$ \\
AS2UDS\ 0266.0  &  $2.745 ^{2.515 }_{3.015}$  & 145  &$0.06\pm0.02$&  $ 0.06\pm 0.00$&  $0.47\pm0.03$&  $ 77\pm  2$&  $0.2\pm0.1$ \\
AS2UDS\ 0269.0  &  $2.545 ^{2.505 }_{2.575}$  & 380  &$0.05\pm0.01$&  $ 0.07\pm 0.00$&  $0.58\pm0.02$&  $113\pm  3$&  $1.0\pm0.2$ \\
AS2UDS\ 0272.0  &  $1.775 ^{1.562 }_{1.988}$  & 380  &$0.08\pm0.01$&  $ 0.12\pm 0.01$&  $0.46\pm0.03$&  $135\pm  2$&  $0.7\pm0.1$ \\
AS2UDS\ 0278.0  &  $2.495 ^{2.435 }_{2.545}$  & 316  &$0.06\pm0.01$&  $ 0.06\pm 0.00$&  $0.70\pm0.02$&  $128\pm  4$&  $0.2\pm0.2$ \\
AS2UDS\ 0280.0  &  $2.685 ^{2.575 }_{2.885}$  & 646  &$0.06\pm0.02$&  $ 0.12\pm 0.01$&  $0.52\pm0.04$&  $ 83\pm  4$&  $2.1\pm0.3$ \\
AS2UDS\ 0286.0  &  $3.235 ^{3.085 }_{3.395}$  & 339  &$0.09\pm0.02$&  $ 0.12\pm 0.01$&  $0.72\pm0.03$&  $ 17\pm  5$&  $1.7\pm0.3$ \\
AS2UDS\ 0297.0  &  $1.675 ^{1.474 }_{1.876}$  & 316  &$0.09\pm0.02$&  $ 0.11\pm 0.01$&  $0.67\pm0.03$&  $ 55\pm  4$&  $0.7\pm0.1$ \\
AS2UDS\ 0298.0  &  $2.475 ^{2.335 }_{2.555}$  & 209  &$0.07\pm0.02$&  $ 0.24\pm 0.02$&  $0.40\pm0.04$&  $ 82\pm  2$&  $4.0\pm0.9$ \\
AS2UDS\ 0302.0  &  $3.665 ^{3.055 }_{4.505}$  & 316  &$0.08\pm0.02$&  $ 0.11\pm 0.01$&  $0.56\pm0.03$&  $ 88\pm  2$&  $1.3\pm0.2$ \\
AS2UDS\ 0306.0  &  $1.535 ^{1.505 }_{1.585}$  &  41  &$0.08\pm0.02$&  $ 0.13\pm 0.01$&  $0.62\pm0.04$&  $ 80\pm  4$&  $2.3\pm0.5$ \\
AS2UDS\ 0310.0  &  $3.305 ^{2.835 }_{4.125}$  & 240  &$0.05\pm0.02$&  $ 0.07\pm 0.01$&  $0.54\pm0.04$&  $ 82\pm  4$&  $0.2\pm0.2$ \\
AS2UDS\ 0315.0  &  $3.305 ^{2.945 }_{4.505}$  & 339  &$0.10\pm0.02$&  $ 0.25\pm 0.01$&  $0.75\pm0.03$&  $167\pm  5$&  $3.8\pm0.7$ \\
AS2UDS\ 0316.0  &  $3.395 ^{3.235 }_{3.565}$  & 646  &$0.10\pm0.02$&  $ 0.20\pm 0.01$&  $0.53\pm0.05$&  $ 36\pm  3$&  $2.9\pm3.0$ \\
AS2UDS\ 0321.0  &  $2.765 ^{2.665 }_{2.805}$  & 955  &$0.07\pm0.02$&  $ 0.10\pm 0.01$&  $0.59\pm0.03$&  $ 12\pm  3$&  $0.3\pm0.1$ \\
AS2UDS\ 0325.0  &  $3.425 ^{3.085 }_{4.115}$  & 490  &$0.07\pm0.02$&  $ 0.10\pm 0.01$&  $0.46\pm0.04$&  $ 74\pm  2$&  $0.9\pm0.2$ \\
AS2UDS\ 0331.0  &  $2.455 ^{2.155 }_{2.715}$  & 219  &$0.05\pm0.01$&  $ 0.05\pm 0.00$&  $0.88\pm0.02$&  $133\pm 23$&  $2.3\pm0.9$ \\
AS2UDS\ 0336.0  &  $5.185 ^{3.025 }_{5.525}$  & 724  &$0.08\pm0.02$&  $ 0.11\pm 0.01$&  $0.66\pm0.03$&  $119\pm  3$&  $1.3\pm0.2$ \\
AS2UDS\ 0343.0  &  $3.275 ^{2.665 }_{4.545}$  & 331  &$0.08\pm0.01$&  $ 0.27\pm 0.01$&  $0.57\pm0.02$&  $ 64\pm  2$&  $5.2\pm0.9$ \\
AS2UDS\ 0347.0  &  $2.655 ^{1.895 }_{3.515}$  & 331  &$0.11\pm0.02$&  $ 0.14\pm 0.01$&  $0.64\pm0.04$&  $ 51\pm  3$&  $1.1\pm0.1$ \\
AS2UDS\ 0348.0  &  $3.405 ^{3.235 }_{3.545}$  & 427  &$0.07\pm0.02$&  $ 0.09\pm 0.00$&  $0.80\pm0.02$&  $ 73\pm  8$&  $3.6\pm0.9$ \\
AS2UDS\ 0353.0  &  $2.625 ^{2.565 }_{2.685}$  & 204  &$0.06\pm0.02$&  $ 0.10\pm 0.01$&  $0.41\pm0.04$&  $137\pm  2$&  $2.0\pm0.4$ \\
AS2UDS\ 0368.0  &  $3.735 ^{3.245 }_{4.215}$  & 617  &$0.08\pm0.01$&  $ 0.11\pm 0.01$&  $0.54\pm0.03$&  $ 43\pm  2$&  $1.5\pm0.2$ \\
AS2UDS\ 0374.0  &  $2.785 ^{2.545 }_{2.965}$  & 372  &$0.09\pm0.01$&  $ 0.11\pm 0.00$&  $0.70\pm0.03$&  $  3\pm  3$&  $0.9\pm0.1$ \\
AS2UDS\ 0389.0  &  $2.615 ^{2.395 }_{2.995}$  & 575  &$0.06\pm0.01$&  $ 0.11\pm 0.01$&  $0.41\pm0.03$&  $ 12\pm  2$&  $1.8\pm0.2$ \\
AS2UDS\ 0395.0  &  $2.345 ^{2.064 }_{2.626}$  & 204  &$0.06\pm0.01$&  $ 0.07\pm 0.00$&  $0.78\pm0.02$&  $ 30\pm  6$&  $1.0\pm0.2$ \\
AS2UDS\ 0402.0  &  $2.575 ^{2.375 }_{2.885}$  & 186  &$0.06\pm0.02$&  $ 0.07\pm 0.01$&  $0.89\pm0.03$&  $ 14\pm 18$&  $0.9\pm0.3$ \\
AS2UDS\ 0403.0  &  $3.015 ^{2.815 }_{3.145}$  & 316  &$0.08\pm0.02$&  $ 0.10\pm 0.01$&  $0.57\pm0.03$&  $103\pm  3$&  $0.3\pm0.1$ \\
AS2UDS\ 0413.0  &  $1.635 ^{1.615 }_{1.655}$  & 204  &$0.07\pm0.02$&  $ 0.09\pm 0.01$&  $0.54\pm0.03$&  $151\pm  3$&  $1.0\pm0.2$ \\
AS2UDS\ 0432.0  &  $2.515 ^{2.375 }_{2.885}$  & 309  &$0.04\pm0.01$&  $ 0.07\pm 0.00$&  $0.39\pm0.03$&  $ 39\pm  2$&  $1.0\pm0.2$ \\
AS2UDS\ 0438.0  &  $3.445 ^{2.995 }_{4.785}$  & 347  &$0.09\pm0.02$&  $ 0.12\pm 0.01$&  $0.47\pm0.05$&  $ 43\pm  2$&  $0.4\pm0.1$ \\
AS2UDS\ 0439.0  &  $3.865 ^{3.165 }_{5.505}$  & 347  &$0.09\pm0.02$&  $ 0.09\pm 0.01$&  $0.95\pm0.03$&  $131\pm 27$&  $1.1\pm0.2$ \\
AS2UDS\ 0440.0  &  $1.985 ^{1.815 }_{2.055}$  & 195  &$0.12\pm0.02$&  $ 0.14\pm 0.01$&  $0.78\pm0.03$&  $137\pm  4$&  $0.8\pm0.1$ \\
AS2UDS\ 0444.0  &  $2.555 ^{2.325 }_{2.735}$  & 145  &$0.08\pm0.02$&  $ 0.10\pm 0.01$&  $0.73\pm0.03$&  $165\pm  5$&  $1.2\pm0.2$ \\
AS2UDS\ 0447.0  &  $1.725 ^{1.705 }_{1.835}$  & 407  &$0.10\pm0.02$&  $ 0.10\pm 0.01$&  $0.92\pm0.03$&  $ 72\pm 21$&  $1.1\pm0.2$ \\
AS2UDS\ 0454.0  &  $3.525 ^{3.005 }_{4.315}$  & 288  &$0.04\pm0.01$&  $ 0.04\pm 0.00$&  $0.49\pm0.02$&  $ 86\pm  3$&  $0.2\pm0.2$ \\
AS2UDS\ 0462.0  &  $3.375 ^{3.035 }_{5.425}$  & 372  &$0.12\pm0.03$&  $ 0.23\pm 0.01$&  $0.61\pm0.05$&  $ 97\pm  3$&  $2.4\pm0.3$ \\
AS2UDS\ 0465.0  &  $2.375 ^{2.335 }_{2.395}$  & 676  &$0.07\pm0.01$&  $ 0.07\pm 0.00$&  $0.92\pm0.02$&  $143\pm 14$&  $1.4\pm0.2$ \\
AS2UDS\ 0470.0  &  $3.275 ^{2.975 }_{3.565}$  & 562  &$0.06\pm0.01$&  $ 0.05\pm 0.00$&  $1.00\pm0.02$&  $144\pm***$&  $0.8\pm0.3$ \\
AS2UDS\ 0481.0  &  $3.025 ^{2.915 }_{3.175}$  & 145  &$0.13\pm0.02$&  $ 0.14\pm 0.01$&  $0.84\pm0.04$&  $ 91\pm  9$&  $1.1\pm0.1$ \\
AS2UDS\ 0483.0  &  $1.855 ^{1.525 }_{2.185}$  & 166  &$0.06\pm0.02$&  $ 0.12\pm 0.01$&  $0.50\pm0.04$&  $161\pm  3$&  $2.9\pm0.8$ \\
AS2UDS\ 0484.0  &  $2.845 ^{2.175 }_{3.635}$  & 398  &$0.04\pm0.01$&  $ 0.07\pm 0.00$&  $0.35\pm0.03$&  $166\pm  2$&  $1.2\pm0.2$ \\
AS2UDS\ 0487.0  &  $3.405 ^{3.275 }_{3.575}$  & 603  &$0.05\pm0.02$&  $ 0.06\pm 0.01$&  $0.50\pm0.04$&  $169\pm  3$&  $0.2\pm0.1$ \\
AS2UDS\ 0489.0  &  $2.205 ^{1.555 }_{2.465}$  & 331  &$0.06\pm0.02$&  $ 0.11\pm 0.01$&  $0.43\pm0.05$&  $132\pm  3$&  $2.2\pm0.4$ \\
AS2UDS\ 0494.0  &  $2.485 ^{2.005 }_{2.965}$  & 182  &$0.07\pm0.02$&  $ 0.11\pm 0.01$&  $0.34\pm0.05$&  $120\pm  1$&  $0.7\pm0.1$ \\
AS2UDS\ 0506.0  &  $1.805 ^{1.555 }_{2.095}$  & 123  &$0.06\pm0.02$&  $ 0.08\pm 0.01$&  $0.69\pm0.03$&  $109\pm  5$&  $0.3\pm0.1$ \\
AS2UDS\ 0513.0  &  $2.285 ^{2.025 }_{2.515}$  &  91  &$0.03\pm0.01$&  $ 0.06\pm 0.00$&  $0.26\pm0.03$&  $ 62\pm  2$&  $1.0\pm0.2$ \\
AS2UDS\ 0521.0  &  $2.955 ^{2.565 }_{3.225}$  & 331  &$0.06\pm0.02$&  $ 0.06\pm 0.01$&  $0.90\pm0.03$&  $ 60\pm 20$&  $0.2\pm0.3$ \\
AS2UDS\ 0531.0  &  $2.765 ^{2.515 }_{2.945}$  & 407  &$0.06\pm0.02$&  $ 0.08\pm 0.00$&  $0.60\pm0.03$&  $158\pm  3$&  $0.5\pm0.1$ \\
AS2UDS\ 0536.0  &  $2.215 ^{1.949 }_{2.481}$  & 269  &$0.09\pm0.02$&  $ 0.15\pm 0.01$&  $0.36\pm0.06$&  $166\pm  1$&  $1.1\pm0.1$ \\
AS2UDS\ 0537.0  &  $2.425 ^{1.795 }_{3.035}$  & 372  &$0.10\pm0.02$&  $ 0.11\pm 0.01$&  $0.77\pm0.04$&  $129\pm  6$&  $0.3\pm0.1$ \\
\hline
\hline
\captionsetup{width=.9\textwidth}
\caption{{\small The parameters for the $870\,\mu$m continuum maps measured using \sersic\ with a fixed $n=1$ and a free Sersic fit in the image plane.
Column 2: photometric redshifts from {Dudzevi{\v c}i{\= u}t{\.e}} et al. (submitted) with the upper and lower limits of the redshifts given in the power and subscript,
Column 3: The star-formation rate (SFR) from {Dudzevi{\v c}i{\= u}t{\.e}} et al. (submitted),
Column 4: The circularised effective radius from the fixed $n=1$ \sersic\ fit,
Column 5: The circularised effective radius from the Free \sersic\ fit,
Column 6: The axis ratio from the fixed $n=1$ fit,
Column 7: The position angle from the fixed $n=1$ fit,
Column 8: The \sersic\ fit from the free \sersic\ fit.}}
\label{table:data}
\end{longtable}

\twocolumn

%\begin{thebibliography}{99}
%\bibliography{AS2UDS}
%\end{thebibliography}

%%%%%%%%%%%%%%%%%%%%%%%%%%%%%%%%%%%%%%%%%%%%%%%%%%

%%%%%%%%%%%%%%%%% APPENDICES %%%%%%%%%%%%%%%%%%%%%

%\appendix


%%%%%%%%%%%%%%%%%%%%%%%%%%%%%%%%%%%%%%%%%%%%%%%%%%

% Don't change these lines
\bsp	% typesetting comment
\label{lastpage}
\end{document}